\documentclass[12pt]{iopart}

\usepackage{graphicx}
\usepackage{siunitx}
\usepackage{pdfpages}
\usepackage{amssymb}
\usepackage[nolist,nohyperlinks]{acronym}
\usepackage{booktabs}
\usepackage{cite} % for combining references

\begin{acronym}
\acro{3D}{three-dimensional}
\acro{PCB}{printed circuit board}
\acro{SOI}{silicon on insulator}
\acro{MEMS}{micro-electro-mechanical system}
\acro{CCD}{charge-coupled device}
\acro{CMOS}{complementary metal-oxide-semiconductor}
\acro{NA}{numerical aperture}
\acro{DC}{direct current}
\acro{AC}{alternating current}
\acro{RF}{radio-frequency}
\acro{UV}{ultraviolet}
\acro{DRIE}{deep reactive ion etching}
\acro{SEM}{scanning electron microscope}
\acro{SAM}{scanning acoustic microscopy}
\acro{imox}{inter-metal oxide}
\acro{OVC}{outer vacuum chamber}
\end{acronym}

\begin{document}

% Mass-Producible Microfabricated 3D Ion Trap with High Trap Depth
% A Microfabricated Ion Trap with 3D Structures
% A Mass-Producible, Microfabricated 3D Ion Trap
\title[Industrially Microfabricated Ion Trap]{Industrially Microfabricated Ion Trap with 1\,eV Trap Depth} % MEMS

\author{S. Auchter$^{1,2,3}$, C. Axline$^{1,4}$, C. Decaroli$^4$, M. Valentini$^2$, L. Purwin$^3$, R. Oswald$^4$, R. Matt$^4$, E. Aschauer$^3$, Y. Colombe$^3$, P. Holz$^5$, T. Monz$^{2,5}$, R. Blatt$^{2,5,6}$, P. Schindler$^2$, C. Rössler$^3$, and J. Home$^4$}

\address{$^1$ These authors contributed equally to this work.}

\address{$^2$ Institut f\"ur Experimentalphysik, Universit\"at Innsbruck, Technikerstra\ss e 25, A-6020 Innsbruck, Austria}

\address{$^3$ Infineon Technologies Austria AG, Siemensstra\ss e 2, A-9500 Villach, Austria}

\address{$^4$ Institut f\"ur Quantenelektronik, ETH Z\"urich, Otto-Stern-Weg 1, CH-8093 Z\"urich,
Switzerland}

\address{$^5$ Alpine Quantum Technologies GmbH, Technikerstra\ss e 17/1, A-6020 Innsbruck, Austria}

\address{$^6 $ Institut f\"ur Quantenoptik und Quanteninformation, \"Osterreichische Akademie der Wissenschaften, Technikerstra\ss e 21\,A, A-6020 Innsbruck, Austria}

\ead{silke.auchter@infineon.com}
\vspace{10pt}
\begin{indented}
\item[]February 2022
\end{indented}

\begin{abstract}

% To solve problems with broad applicability, future quantum computers will demand control over an enormous number of qubits, and microfabricated ion traps can help achieve this.
% However, planar traps will ultimately require robust, industrial-scale fabrication compatible with innovations like integrated photonics and electronics, and they will benefit from trap depths and low heating rates generally provided by three-dimensional traps.
% We present a 3D ion trap fabricated on 8-inch wafers in a large-scale \acs{MEMS} microfabrication process that provides identical traps with high yield.
% Electrodes are patterned on the surfaces of two opposing wafers bonded to a spacer.
% We design a trap depth of 1\,eV for ${}^{40}$Ca${}^{+}$ ions and measure device characteristics in good agreement.
% We also characterize and compensate stray electric fields at multiple trapping sites, and find motional heating rates \textless0.04\,phonons/ms (at 1\,MHz, 185\,K, and \SI{200}{\micro\meter} surface-ion distance) that improve significantly at cryogenic temperatures.
% This fabrication approach represents a significant advance towards scalable, fault-tolerant quantum computing with trapped ions.

%% Jonathan's rewrite: (he also says, skip the first two sentences for short-format journals)
Scaling trapped-ion quantum computing will require robust trapping of at least hundreds of ions over long periods, while increasing the complexity and functionality of the trap itself.
Symmetric \ac{3D} structures enable high trap depth, but microfabrication techniques are generally better suited to planar structures that produce less ideal conditions for trapping.
We present an ion trap fabricated on stacked 8-inch wafers in a large-scale \acs{MEMS} microfabrication process that provides reproducible traps at a large volume.
Electrodes are patterned on the surfaces of two opposing wafers bonded to a spacer, forming a \ac{3D} structure with $\SI{2.5}{\micro\meter}$ standard deviation in alignment across the stack.
We implement a design achieving a trap depth of 1\,eV for a ${}^{40}$Ca${}^{+}$ ion held at \SI{200}{\micro\meter} from either electrode plane.
We characterize traps, achieving measurement agreement with simulations to within $\pm 5$\% for mode frequencies spanning 0.6--3.8\,MHz, and evaluate stray electric field across multiple trapping sites. % this was worded weirdly before
We measure motional heating rates over an extensive range of trap frequencies, and temperatures, observing 40\,phonons/s at 1\,MHz and 185\,K.
This fabrication method provides a highly scalable approach for producing a new generation of \ac{3D} ion traps.

\end{abstract}
\acresetall

%
%Uncomment for keywords
\vspace{2pc}

% submitto{\JPA}
% maketitle

\section{Introduction}

%% Jonathan's rewrite:
Quantum computing \cite{Shor1994,Grover1996} and measurement \cite{Wineland2002} constitute areas in which quantum systems can provide an advantage relative to classical devices.
However, this advantage generally only becomes apparent once a sufficient system size is achieved, which presents a challenge to all technologies being used to pursue this advantage. % quantum speedup.

In the context of quantum computing, trapped ion systems represent a promising approach that fulfills the DiVincenzo requirements \cite{DiVincenzo1995,DiVincenzo2000}.
Many of the most significant results in the field of trapped-ion quantum computing have been achieved using macroscopic linear traps, which hold tens of ions in a single potential well \cite{Lechner2016,Pagano2018,Friis2018,Joshi2020}.
In such systems, high fidelity qubit operations \cite{Srinivas2021}, long coherence times \cite{Wang2021}, and control over long ion strings with about 50~qubits \cite{Pagano2018,Kranzl2021} have been demonstrated.
However, in order to scale to systems with enough resources to suppress errors using error correction, a modular approach based on inter-connected sub-units will likely be required.

Two modular approaches are under consideration, and are promising for scaling to more than 100~ions.
In the first approach \cite{Cirac1997, Monroe2014}, separated ion trap modules are interfaced by establishing entanglement through photonic links, an approach that will eventually place strong demands on the reproducibility of traps that must operate reliably in many different setups. %\cite{Monroe2014}
% The second approach, which guides this work, has multiple trapping zones situated in a single trap structure \cite{Kielpinski2002,Holz2020}, where zones are interfaced by shuttling ions between them using voltages applied to segmented electrodes \cite{Walther2012,Bowler2012}.
The second approach uses multiple trapping zones situated in a single trap structure \cite{Kielpinski2002,Wan2019,Pino2021,Clercq2016,Holz2020}, where zones are interfaced by shuttling ions between them using voltages applied to segmented electrodes \cite{Walther2012,Bowler2012,Kaushal2020}.
Scaling up using these approaches has been challenging due to its complexity: the realization of precisely fabricated trap structures with large numbers of electrodes, and the task of wiring these up and controlling them.
 
The use of well-established microfabrication techniques \cite{Hughes2011}, including those using \ac{MEMS} technology \cite{Cho2015,Blain2021}, is very attractive to realize traps that meet the demands of complexity \emph{and} reproducibility that are critical to both approaches. % good, MEMS references are both to reviews, seems appropriate
However, while deposition and structuring of patterned conducting and insulating layers on a surface is highly accurate and can achieve high levels of complexity (as for example in ref. \cite{Hughes2011,Britton2009,Amini2010,Seidelin2006,Chiaverini2005,Kielpinski2002,Holz2020}), these techniques are generally ill-suited to realize \ac{3D} structures extending beyond several \SI{}{\micro\meter} in the out-of-plane direction.
Therefore, many early investigations were constrained to a planar electrode geometry, which results in lower trap depth (typically $\sim$ 100\,meV) and highly asymmetric field lines \cite{Chiaverini2005,Seidelin2006}, complicating the control of traps and making them more susceptible to ion loss \cite{Doret2012,Wright2013}.
Demonstrations of multi-segment \ac{3D} trap electrode structures have been achieved by stacking multiple layers of laser machined or etched material \cite{Hensinger2006,Blakestad2009,Kaushal2020,Decaroli2021a}, however the methods used to produce these traps were not compatible with standard semiconductor fabrication.
However, these have achieved trap depths above 1\,eV which results in long ion storage times \cite{Pagano2018,Bruzewicz2019}, and have demonstrated higher levels of control in advanced tasks such as junction transport \cite{Blakestad2009} and non-adiabatic ion transport \cite{Walther2012, Bowler2012}.

In this article, we describe the design, fabrication, and characterization of a microfabricated ion trap that, in contrast to monolithic microfabricated traps \cite{Wilpers2012}, combines surface-electrode structures on multiple, precisely aligned wafers that results in a \ac{3D} structure.
This achieves a trap depth that surpasses the typical depth present in microfabricated surface-electrode ion traps by one order of magnitude.
% Fabrication uses an extensible \ac{MEMS} process on an industrial fabrication line, which realizes high-volume, reproducible production.
Fabrication uses a \ac{MEMS} process on an industrial fabrication line, which realizes reproducible production that can be easily adapted to incorporate new designs. %high-volume, 
% While \ac{MEMS} technology has previously been successfully used to produce planar ion traps \cite{Cho2015,Blain2021} with trap electrodes fabricated in layers on a single surface, 
Our fabrication process is streamlined and suited for mass production, including assembly and packaging.

% We present the design concept of the trap and details of trap fabrication.
We characterize ion trapping performance by trapping ${}^{40}$Ca${}^{+}$ ions in a cryogenic apparatus that allows extensive variation of the trapping conditions. % ... like voltages, temperatures, ...
We compare measured trap parameters with the simulated model, establishing broad agreement over motional mode frequencies between 0.6\,MHz and 3.8\,MHz.
Studies are also carried out over temperatures between 75\,K and 300\,K.
% By comparing measured and simulated trap parameters, we infer that the trapping depth reaches the design goal of at least 1\,eV.
% We measure motional heating rates close to the state of the art \cite{Brown2021} and sufficient for multi-qubit operations in a quantum processor or measurement device, and characterize stray static electric and magnetic field components stemming from the trapping fields.
We measure motional heating rates, finding them comparable to traps with ion-surface distance close to our \SI{200}{\micro\meter} value \cite{Brown2021}, and characterize stray static-electric and magnetic field components.
We thus provide evidence that this \ac{MEMS} technology is suitable for realizing ion traps for scaled-up quantum technologies.

\section{Ion Trap Design and Fabrication}

\subsection{Trap concept and design}
\label{sec:concept}

\begin{figure}[htb]
    \centering
    \includegraphics[width=1\textwidth]{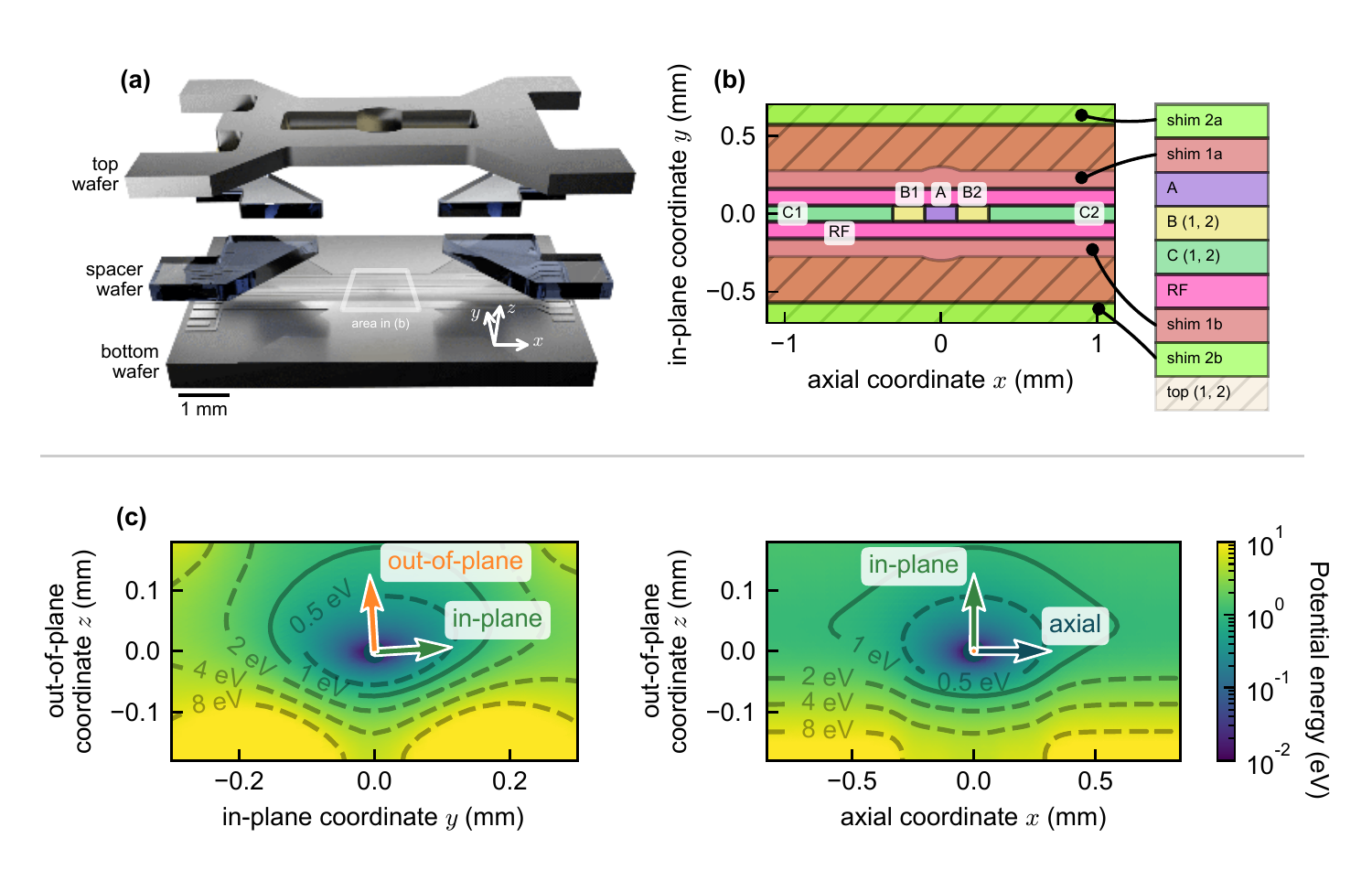}
    \caption{Trap concept.
    a) Exploded view showing the three wafers that form the ion trap.
    The bottom wafer includes metal \ac{DC} and \ac{RF} electrodes.
    The glass spacer defines a nominal distance of $\SI{400}{\micro\meter}$ between top and bottom electrodes.
    Voltages applied to electrodes on the top wafer can adjust the confining potential.
    Laser access is possible from all four sides of the trap, and a slit between the top electrodes enables fluorescence detection from the top.
    b) Top (hatched) and bottom (filled) wafer electrode geometry in the region near the central trapping zone.
    The symmetric electrodes `B1', `B2', `C1', `C2', `top 1', and `top 2' are designed to be controlled independently. % I'm stepping carefully here, because with these two particular traps we can't control the top independently...
    c) Simulated trap potential energy and motional mode vectors for radial (left) and axial (right) cross sections through the center trapping site for the voltage set resulting in a trap depth of 1\,eV (given in Table~\ref{tab:voltages}).
    The trap depth exceeds 1\,eV as indicated by 1\,eV contour levels (solid lines) that fully enclose the trapping site.
    Due to voltage limits (Section \ref{sec:waferbond}), this set only provides a 3.6$^{\circ}$ rotation of the radial modes about the trap ($x$) axis.
    }
    \label{fig:concept}
\end{figure}

% Here, we describe the features and functionality of 
Our microfabricated \ac{3D} ion trap design, shown in Figure~\ref{fig:concept}a, comprises three wafers.
A bottom wafer carries \ac{DC} and \ac{RF} signals on patterned segmented electrodes. %, following the multi-metal-layer technology used to produce a previous trap \cite{Holz2020}.
% The trap is three-dimensional because it contains structures extending far out of the surface electrode plane, but it also is 'electrically' 3D because electrodes are on both wafers...
Electrodes on the top wafer electrically extend the trap into the third dimension, creating a \ac{3D} structured trap. % imparting its \ac{3D} nature.
Our design serves as a proof of concept for a future \ac{3D} linear trap with \ac{RF} electrodes on top and bottom wafers.
Applying \ac{RF} voltage to both bottom and top wafer allows for a more harmonic potential compared to surface-electrode traps and increases the intrinsic \ac{RF} trap depth.
As a first step towards this goal we incorporate \ac{DC} electrodes on the top wafer to adjust the potential, including by redistributing confinement between the radial directions \cite{Wesenberg2008} to improve trap depth. %incorporating \ac{DC} electrodes on the top wafer to adjust the potential, including by redistributing confinement between the radial directions \cite{Wesenberg2008} to improve trap depth.
% providing the additional confinement for enhanced trap depth.
% Electrodes on the top wafer electrically extend the trap into the third dimension.
% Critically, this provides [an RF ground, DC electrodes, opportunity to add RF] which affects confinement [in some way]
A spacer wafer connects bottom and top wafer, and provides optical access.
Such an electrode configuration is similar to previously realized surface traps with an additional metal lid on top to enhance trap depth \cite{Brandl2016a,Kumph2016a}, but the presence of a patterned top wafer allows for extensibility and more complex control over the potential.

Electrostatic finite element method simulations were used to design the \ac{DC} and \ac{RF} electrode geometry shown in Figure~\ref{fig:concept}b. % that supported a trap depth of at least 1\,eV.
The simulations consider the full potential $\Phi$ in which a ${}^{40}\rm{Ca}^{+}$ ion is trapped (over spatial coordinates indexed by $\textbf{r}$),
\begin{equation}
    \Phi({\textbf{r}}) = \Phi_{\rm{dc}}({\textbf{r}}) + \bar{\Phi}_{\rm{rf}}({\textbf{r}})
\end{equation}
that includes the \ac{DC} potential $\Phi_{\rm{dc}}$ and \ac{RF} potential $\bar{\Phi}_{\rm{rf}}$ in the pseudopotential approximation, wherein the force is time-averaged over an oscillation period of the \ac{RF} drive \cite{Kienzler2015}.
The pseudopotential is given by 
\begin{equation}
    \bar{\Phi}_{\rm{rf}} = e \frac{|\nabla \Phi_{\rm{rf}}(\textbf{r})|^2}{4 m \Omega_{\rm{rf}}^2}
\end{equation}
where $e$ is the elementary charge, $m$ is the atomic mass, and % $\Phi_{\rm{rf}}$ is defined by % shortened
\begin{equation}
    \Phi_{\rm{rf}}(y, z) = \frac{1}{2} V_{\rm{rf}} \cos(\Omega_{\rm{rf}} t) (\alpha y^2 - \beta z^2)
\end{equation}
for \ac{RF} amplitude $V_{\rm{rf}}$, \ac{RF} frequency $\Omega_{\rm{rf}}$, and geometry-dependent factors $\alpha = \beta$ adhering to Laplace's equation for a linear Paul trap.
In a simulation of the potential energy landscape, the trap depth $D$ is defined as the energy $\Phi({\textbf{r}_s}) - \Phi(0)$ at the lowest saddle point $\textbf{r}_s$ in any direction away from the central trapping site at $\textbf{r} = 0$.
If the ion's energy exceeds this depth, it may escape the trap.

% The trap was designed to produce a 1\,eV-deep pseudopotential at $\Omega_{\rm{rf}} /(2 \pi) \sim \SI{40}{\mega\hertz}$ and $V_{\rm{rf}} \sim 250$\,V, but in the experiments here we opted to use 183\,V at $\Omega_{\rm{RF}}/(2 \pi) \sim 20.6$\,MHz to accommodate an available resonator while reducing \ac{RF} dissipation and trap temperature (see Section \ref{sec:expt}).
% To achieve the same trap depth at lower $V_{\rm{rf}}$, the \ac{DC} part of the potential was changed \cite{Wesenberg2008}.
The trap typically operates at $V_{\rm{rf}}=183$\,V and $\Omega_{\rm{RF}}/(2 \pi) \sim 20.6$\,MHz. % to accommodate an available resonator while reducing \ac{RF} dissipation and trap temperature (see Section \ref{sec:expt}).
% To achieve the same trap depth at lower $V_{\rm{rf}}$, the \ac{DC} part of the potential was changed \cite{Wesenberg2008}
The overall trap depth was increased by biasing \ac{DC} voltages so as to increase confinement into the out-of-plane direction ($\hat{z}$) at the expense of confinement in the in-plane direction ($\hat{y}$) \cite{Wesenberg2008}, which does not generally limit trap depth in this design.
The resulting voltage set solution, given in Table~\ref{tab:voltages}, produces the potential shown in Figure~\ref{fig:concept}c, which remains highly harmonic around the trapping site (see \ref{sec:anharmonicity}).

This potential gives a trap depth that is limited equally by 1.1\,eV saddle points near to the $x$ and $z$ axes (at distances of \SI{860}{\micro\meter} and \SI{180}{\micro\meter} from the trap center, respectively), while the lowest potential barrier along the $\hat{y}$ direction is 2.5\,eV high around \SI{320}{\micro\meter} away.
Higher trap depths could be reached if voltages were increased, ultimately limited by dielectric breakdown voltages between metal layers.
In traps using a comparable layer stack-up and electrode spacing, breakdown voltages exceeding \SI{800}{\volt} were measured \cite{Holz2020}.

The quadrupole confinement produced by this trap configuration can be compared to an ideal quadrupole potential (which has depth $D = q V_{\rm{rf}} / 4$ for dimensionless Mathieu parameter $q$, using the conventions in ref. \cite{Douglas2015}) to extract a depth-parameterized trap efficiency $\eta = 4 D / (q V_{\rm{rf}})$ in terms of intrinsic (\ac{RF}-only) trap depth \cite{Wesenberg2008}.
We simulate the potential of this trap with grounded top electrodes, as well as the planar-fabricated design in ref. \cite{Mehta2020}, finding $\eta \sim 1\%$ in both cases. % (1.1 percent and 1.3 percent, respectively)
This is comparable to 2--5\% efficiencies calculated in other surface-electrode geometries \cite{Wesenberg2008, Schuster2011}.
Similar results are to be expected, since the quadrupole field is generated similarly (in plane).
% Trap depth
Simulations of our microfabricated \ac{3D} trap predict an \emph{effective} trap efficiency $\eta \sim 5\%$ under operational conditions, when \ac{DC} voltages are used to modify the radial potential.
% With minor modifications, \ac{RF} electrodes could be added to the top wafer to produce a more symmetric quadrupole potential, like that of a traditional \ac{3D} trap, increasing intrinsic trap efficiency.
The \emph{intrinsic} trap efficiency can be improved, for example, by applying \ac{RF} drives to electrodes patterned on both top and bottom wafers.
Our simulations show that electrode configurations can be found that produce a quadrupole potential similar to that of traditional \ac{3D} traps \cite{Blakestad2010}, with increased symmetry and trap depth relative to our present arrangement. % can produce a more symmetric quadrupole potential like that of a traditional 3D ion trap, further increasing trap depth and intrinsic trap efficiency.
However, while trap depth or trap efficiency are useful metrics for quantifying and comparing trap properties, additional benefits afforded by a \ac{3D} geometry --- like increased symmetry and harmonicity --- should be considered as well. %!!

% Finite element method simulations of the electric field from \ac{DC} and \ac{RF} electrodes, shown in Figure~\ref{fig:concept}b, were performed initially to aid trap design.
We also used simulation data to estimate mode frequencies, calculate stability parameters, predict the effect of static electric field offsets, and find electrode voltages that produce a specified electric potential.
Convex optimization techniques were used to solve for voltage sets that could independently control axial confinement, rotation of the confining quadrupole of the potential in the radial plane, and micromotion compensation in three directions \cite{Allcock2010}. % radial mode rotation

% The trap was designed to produce a 1\,eV potential at $V_{\rm{rf}} = 250$\,V and $\Omega_{\rm{RF}}/(2 \pi) \sim 20.6$\,MHz, but we opted to use 183\,V to reduce \ac{RF} dissipation and lower trap temperature.

% removed: The resulting voltage set, and the associated electrode geometry, is shown in Figure~\ref{fig:depth}b.
% The resulting voltage set applied to achieve a 1\,eV depth is given in Table~\ref{tab:voltages}. %, associated electrode geometry, and potential energy of the 1\,eV-deep set are shown in Figure~\ref{fig:matching}b--d.

\begin{table}[htb]
\centering
\caption{
    We apply trapping voltages based on two standard sets, with simulated trap depths near 1\,eV and 0.2\,eV, while applying \ac{RF} at $\Omega_{\rm{RF}}/(2 \pi) \sim 20.6$\,MHz.
    Electrodes labels correspond to those in Figure~\ref{fig:concept}b.
    % The \ac{DC} voltages are assembled from parameterized voltage sets; by changing parameters one can adjust axial confinement, radial mode rotation, and micromotion compensation. % PH: this will be obvious % CA: I guess we mention it in the text, so we can leave it out if we must
    The 1\,eV set produces the deep trapping potential displayed in Figure~\ref{fig:concept}c, while the 0.2\,eV set is the starting point for most other measurements in this work.
    The 1\,eV set is scaled up from the 0.2,\,eV set and then further adjusted to increase curvature in the out-of-plane ($\hat{z}$) direction. %  is nearly related to the 0.2\,eV set by a scaling factor
    }
\begin{tabular}{lrr}
\toprule
{} &  1\,eV set (V) &  0.2\,eV set (V) \\
\midrule
shim 2a    &          0.04 &            8.16 \\
shim 1a    &         11.32 &            0.24 \\
A          &          7.66 &           -1.07 \\
B (1, 2)   &          8.95 &            5.29 \\
C (1, 2)   &         17.18 &            4.91 \\
shim 1b    &         12.77 &            1.28 \\
shim 2b    &        -24.03 &           -9.04 \\
top (1, 2) &         10.46 &            0.58 \\
\midrule
$V_{\rm{rf}}$ &     183 &     155\\
\bottomrule
\end{tabular}
\label{tab:voltages}
\end{table}

\subsection{Trap structure}
\label{sec:structure}

\begin{figure}[htb]
    \centering
    \includegraphics[width=1\textwidth]{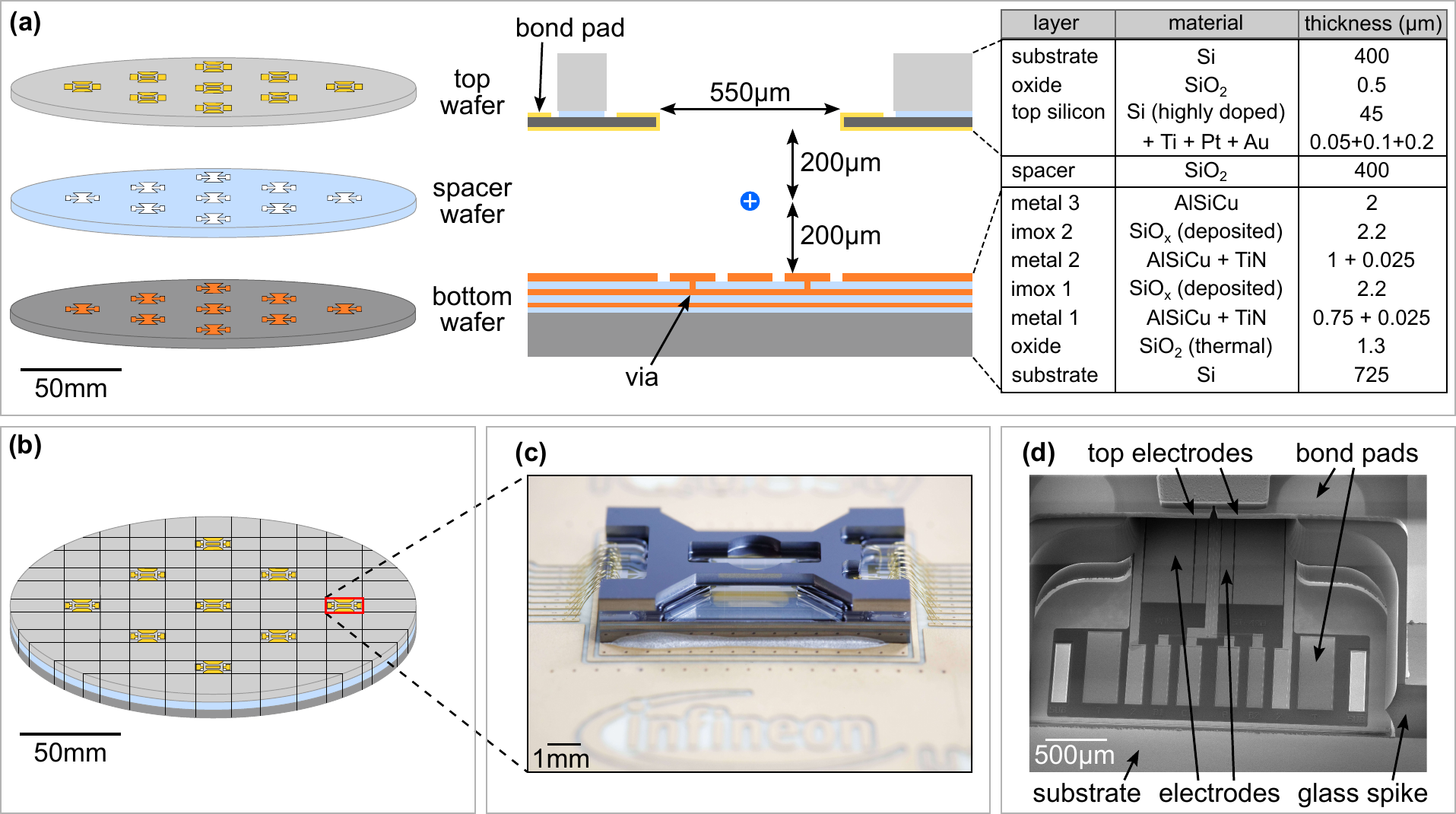}
    \caption{Trap fabrication.
    a) Schematic, radial cross section, and layer stack of the bottom, spacer, and top wafers, which are structured individually.
    Top electrodes are electrically connected by wirebonding to a bond pad on the highly-doped silicon.
    The bottom wafer consists of three metal layers (orange), isolated by oxide (blue).
    Vias in the \acf{imox} connect individual metal layers. 
    Laser access is possible through openings in the spacer wafer, with an NA of 0.75 corresponding with openings away from the axis, and an NA of 0.2 along the axial direction (though access is reduced by wirebonds).
    The table lists all layers with corresponding material and thickness.
    b) The three structured wafers are joined through anodic wafer bonding and the single ion traps are separated by mechanical dicing of the bonded wafer.
    c) An individual chip after dicing, glued and wirebonded to the carrier \ac{PCB} for electrical contact.
    d) A \ac{SEM} image of the ion trap structure at the position of the top electrode bonding pad.
    % The radial displacement of the top electrodes to the bottom electrodes after wafer bonding is measured to be $\SI{7(2)}{\micro\meter}$.
    Analysis of \ac{SEM} images verify a bond alignment accuracy of $\text{3--4}\,\si{\micro\meter}$ for each interface, which together with the lithographic tools' specifications leads to a $\SI{2.5}{\micro\meter}$ standard deviation in alignment tolerance for the full stack.}
    \label{fig:fabrication}
\end{figure}

The structure of the three stacked wafers comprising the \ac{3D} trap, including details on layer materials and thicknesses, is shown in Figure~\ref{fig:fabrication}a.

% bottom geometry
The trap's bottom wafer is based on a $\SI{725}{\micro\meter}$ thick silicon substrate and uses the same multi-metal-layer technology used to produce a previous surface-electrode trap \cite{Holz2020}.
Axial confinement is provided by five segmented \ac{DC} electrodes (`A', `B1', `B2', `C1', `C2') centered on the trap axis (Figure~\ref{fig:concept}b) so that ions can be trapped along the axis at positions within $\pm \SI{250}{\micro\meter}$ of the center.
% They support a single- or a double-well potential (one or two trapping sites).
DC compensation electrodes (`1a' and `1b', which are $\SI{105}{\micro\meter}$ displaced from the $x$-axis and $\SI{100}{\micro\meter}$ wide) on both sides of the trap axis run along the length of the trap and are used for micromotion compensation and radial rotation of the quadrupole moment of the potential. % radial mode tilting.
Two \ac{RF} electrodes ($\SI{360}{\micro\meter}$ displaced from the $x$-axis, $\SI{400}{\micro\meter}$ wide) generate confinement in the radial direction.
These are placed symmetrically along the axis and are connected together at one end of the trap.
The remainder of the trap surface is taken up by \ac{DC} electrodes (`2a', `2b') that can be grounded or used for additional compensation.
% Connecting to electrodes patterned at any location, such as three central islands (each $\SI{200}{\micro\meter}$ long), is possible using vias between the three metal layers, which are electrically separated by inter-metal oxide (imox).
Vias between the three metal layers allow connections to electrodes at any location, such as the three isolated central islands, each $\SI{200}{\micro\meter}$ long.
%These metal layers are made of AlSiCu and are electrically separated by inter-metal oxide (imox) deposited as $\text{SiO}_{x}$.
The lowest metal layer shields the substrate to avoid \ac{RF} loss in the silicon substrate \cite{Krupka2006} and aims to prevent the creation of charge carriers induced by stray laser light \cite{Mehta2014}.

% top geometry
The top wafer consists of $\SI{445}{\micro\meter}$-thick highly-doped \ac{SOI}.
The highly doped silicon has a thickness of $\SI{45}{\micro\meter}$ and a resistivity of $\rho = \text{1--2}\,\si{\milli\ohm \centi\meter}$ at room temperature.
This forms the top layer of the ion trap and incorporates two individually controlled \ac{DC} electrodes.
In this design, no \ac{RF} is present on the top wafer.
The top wafer electrodes are split to create a $\SI{550}{\micro\meter}$-wide slit, providing optical access to the ion with a \acl{NA} \acs{NA} $= 0.75$ that can be used for fluorescence detection.
Two bonding pads on the top wafer allow for electrically connecting the top electrodes to additional bond pads on the bottom wafer or directly to the carrier \acl{PCB}.

% spacer
The spacer wafer is made of $\SI{400}{\micro\meter}$-thick borosilicate glass featuring gaps at the sides and ends through which laser beams can pass above the surface, with an \acs{NA} of 0.75 corresponding with openings away from the axis, and an \acs{NA} of 0.2 along the axial direction. %that allow for laser access within $\pm 55^{\circ}$ of the radial ($yz$) plane and $\pm 10^{\circ}$ of the axial plane for cooling and manipulation of trapped ions.
% technically +/- 10 degrees from the axis and +/- 55 degrees from the y-axis, based on spacer spike positions
% NA of 0.75 in radial and 0.2 in axial direction.
Additionally, wirebond pads at the ends of the trap are spaced to allow axial beam access.
Thus, beams should be able to cross the center of the trap in the $xy$ plane without obstruction at angles up to $\pm 55^{\circ}$ from the $y$-axis, and about $\pm 10^{\circ}$ from the trap ($x$) axis.

\subsection{Trap fabrication}
\label{sec:fabrication}

In order to realize the \ac{3D} trap structure described above, three individual 8-inch wafers were wafer-bonded, as illustrated in Figure~\ref{fig:fabrication}b.
Fabrication of the ion trap was carried out in an industrial cleanroom facility \footnote{Infineon Technologies Austria AG, Villach, Austria}.

The silicon bottom wafer comprises three AlSiCu metal layers created via sputter deposition. These metal layers are isolated by \acf{imox} layers, deposited as silicon oxide.
All metal and oxide layers are structured by optical lithography followed by plasma etching.
Details on the fabrication process for this technology can be found in ref. \cite{Holz2020}.

The $\SI{45}{\micro\meter}$-thick top electrodes and the $\SI{400}{\micro\meter}$-thick undoped silicon substrate of the top wafer are etched via \ac{DRIE}, with an additional recess of the substrate silicon to provide the high \acs{NA} (see the cross section in Figure~\ref{fig:fabrication}a).
To shield the ion from silicon surfaces, the electrodes are gold coated on both sides using a shadow-mask-evaporated stack of $\SI{50}{\nano\meter}$ titanium, $\SI{100}{\nano\meter}$ platinum, and $\SI{200}{\nano\meter}$ gold.
Here, titanium serves as an adhesion layer and platinum as a diffusion barrier between the gold and the silicon substrate.
However, there are areas of exposed substrate silicon ($\rho = \text{1--10}\,\si{\ohm \centi\meter}$) on the top wafer, which, if charged up, could lead to stray electric fields at the ion's position. % in direct line-of-sight to the ion.

The spacer wafer is made of borosilicate glass with a coefficient of thermal expansion matched to silicon to minimize strain at cryogenic temperatures.
All glass structures are created by a wet-chemical etch of $\SI{280}{\micro\meter}$ from front and back sides, giving rise to an $\SI{84}{\micro\meter}$ wide spike on the sidewalls (visible in Figure~\ref{fig:fabrication}d).
% Since the sidewalls are not coated with metal, field noise from charges induced in the dielectric could lead to motional heating \cite{Teller2021}.
One concern this raises for the operation of the trap is that the uncoated glass surfaces could possibly lead to undesired stray charge buildup, causing additional electric fields, or lead to increased motional heating \cite{Teller2021}.
We study this effect in Section \ref{sec:field}.

\subsection{Wafer bond, dicing and assembly}
\label{sec:waferbond}

Anodic wafer bonding \cite{Wallis1969} is a widely used method in industrial packaging and manufacturing, especially in the field of sensor systems \cite{Peeters1992,Rogers2005}.
With its high bonding strength and robustness \cite{He2015}, it provides a suitable technique to connect individual wafers to form a \ac{3D} ion trap.

We developed a two-step anodic bonding process, optimized for inclusion in the industrial fabrication line: first, the bottom wafer is bonded to the glass spacer, and subsequently, the resulting double stack is bonded to the top wafer.
This two-step process allows for semi-automatic optical alignment of both interfaces despite the opacity of silicon wafers that form the outer layers of the triple stack.
We found optimal bond parameters of $\SI{330}{\celsius}$ bond temperature and a bond voltage of $\SI{300}{\volt}$, as verified by \ac{SAM} analysis, die-shear tests, and repeated cryo-cycling between room temperature and $T = \SI{4}{\kelvin}$ (within ca. 5 hours) or $T = \SI{70}{\kelvin}$ (within ca. 1 minute).

The ion traps on the bonded wafer are separated by mechanical dicing as shown in Figure~\ref{fig:fabrication}b. 
Then, each ion trap is glued and electrically connected to a carrier \ac{PCB} as part of an industrial packaging process, based on automated lead-frame handling, semi-automatic die-attach and automated wirebonding, resulting in the final chip shown in Figure~\ref{fig:fabrication}c.

Using \ac{SEM} imaging of the wafer stack cross section (Figure~\ref{fig:fabrication}d) the bond alignment accuracy of $\text{3--4}\,\si{\micro\meter}$ for each interface was verified.
The standard deviation in the alignment tolerance of $\SI{2.5}{\micro\meter}$ across the full wafer stack is in agreement with the lithographic and bonding tools' specifications.
A tilt between top and bottom wafer caused by variations in the glass wafer thickness ($\ll \text{1}\,\si{\micro\meter}$) can be neglected for a single chip.

In the batch of traps characterized in this work, \ac{SEM} imaging revealed a fabrication defect where the two top electrodes were connected together. % in the batch that was used for characterizing the ion trap.
% Thus the radial rotation we could apply while also providing a deep well was limited.
Only the four shim (`1a', `1b', `2a', `2b') and two top (`top 1', `top 2') electrodes have significant radial potential moments that can rotate the quadrupole of the potential.
Without this moment from individual top electrodes, higher voltages had to be applied to the shim electrodes to rotate radial modes, which limited the rotation angle of the modes seen in Figure~\ref{fig:concept}c.
Repeated analysis verified that the defect was corrected in later batches.

Trap fabrication can be done at high volume.
Following introduction of the design into the production line, a one-time run-through of the manufacturing cycle (including packaging) produces 50 identical ion traps from one wafer in 4--6 weeks.
Typical fabrication runs include lots of 25~wafers.

\section{Experimental characterization}
\label{sec:expt}

\begin{figure}[htb]
    \centering
    \includegraphics[width=1.0\textwidth]{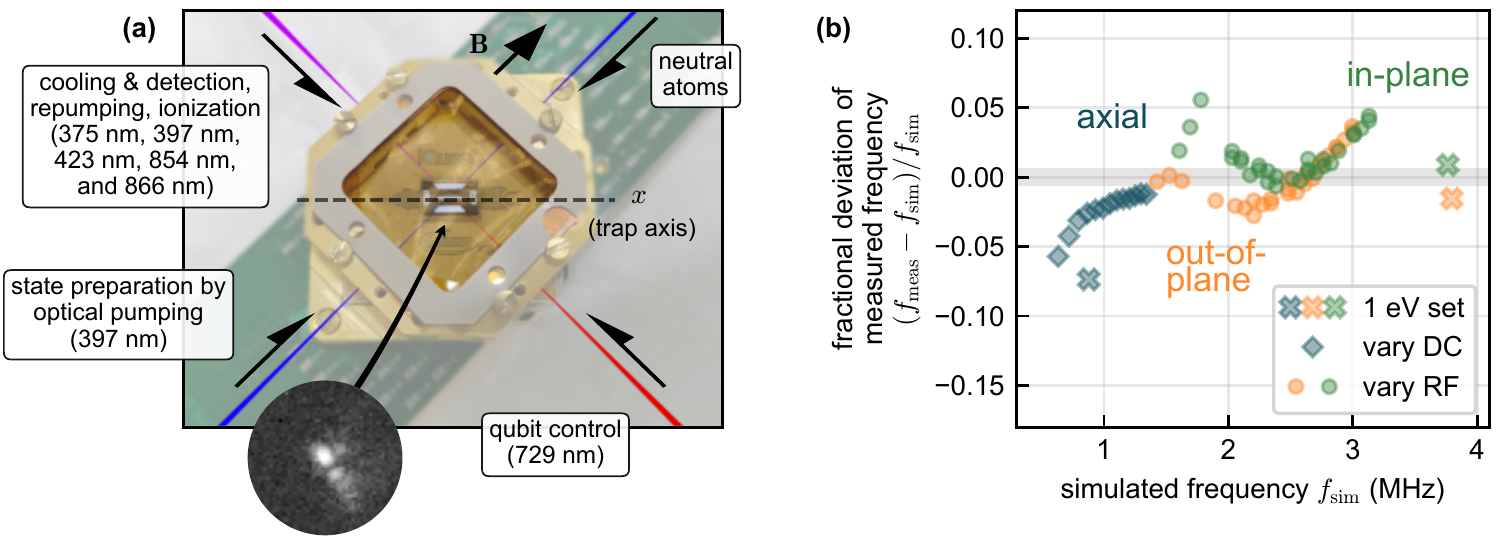}
    \caption{Measurement of motional mode frequencies.
    a) The traps are tested in a cryogenic apparatus.
    Laser beams (purple) used for cooling and detecting (397\,nm), repumping (854\,nm, 866\,nm), and photoionizing calcium atoms (375\,nm, 423\,nm) are introduced across the trap surface at 45$^{\circ}$ to the trap axis and 90$^{\circ}$ to the applied magnetic field $\textbf{B}$.
    Qubit manipulation is done with an opposite-facing beam (red) at 729\,nm.
    Light for state preparation using optical pumping at 397\,nm enters orthogonally (blue), and neutral atomic flux follows an opposing path (while loading ions into the trap).
    b) Comparison of simulated and measured motional mode frequencies in trap \#2.
    Mode frequencies (axial: blue, in-plane radial: green, out-of-plane radial: orange) are found using sideband spectroscopy.
    Axial frequencies are measured in response to changes to \ac{DC} voltages designed to adjust axial confinement (diamonds).
    Radial mode frequencies are measured in response to changes in \ac{RF} amplitude (circles).
    Results for the 1\,eV trap depth voltage set (`X' markers) are also shown.
    % Each point represents a different \ac{DC} or \ac{RF} configuration: 1\,eV trap depth (`X' markers), or swept axial confinement voltages (diamonds) or \ac{RF} amplitudes (circles) for voltage sets designed with lower trap depth.
    % Deviations may come from variable voltage gain and stray field curvature, which may also vary between data sets.
    Measurement uncertainty is smaller than the marker size.}
    \label{fig:matching}   
\end{figure}

% To measure trap characteristics linked to trap depth, and to probe other performance metrics, 
Measurements of traps with ions can validate the simulation, design, and fabrication process.
We test our traps by placing them at the lowest-temperature (``base'') cooling stage of a cryogenic apparatus \cite{Decaroli2021} and trapping ${}^{40}\rm{Ca}^{+}$ ions.
% the trap is attached to a \ac{DC} filter board (with first-order, 36\,kHz-cutoff-frequency low-pass filters) and inserted into a cryogenic apparatus at the lowest-temperature (``base'') stage \cite{Decaroli2021}.
Traps are attached to a \ac{DC} filter board (with first-order, 36\,kHz-cutoff-frequency low-pass filters).
\ac{DC} signals are generated, pass through additional filters at room temperature, and are delivered to the base stage via flexible \ac{PCB} ribbons to the trap filter board.
These ribbons also carry lines from the \ac{DC} voltage source that reference both \ac{DC} and \ac{RF} ground on the trap.
The \ac{RF} signal is delivered via coaxial cables and stepped up using a helical coil resonator (anchored to the same temperature stage, close to the trap), which sets the preferred \ac{RF} drive frequency. % reference for HC resonator? or rely on Chiara ref.
The magnitude $V_{\rm{rf}}$ was inferred using a rectifier on the filter board \ac{PCB} (\ref{sec:rfcal}).
A magnetic field $|\textbf{B}| \sim 5$\,G provided by external coils splits the energy levels of trapped ions using the Zeeman effect and defines a quantization axis.
Lasers for photoionization, detection, cooling, state preparation, and qubit manipulation enter the trap structure along the directions as shown in Figure~\ref{fig:matching}a.

This work describes two traps (\#1 and \#2) from the same fabrication batch tested sequentially in the cryostat, which was operated at a base cooling stage temperature $T_{\rm{base}} \sim 6.5$\,K.
The first trap was attached to a \ac{PCB} that was directly thermally anchored to the base cooling stage.
The second trap was placed on a \ac{PCB} that included a calibrated Pt1000 thermistor, and was more weakly thermally linked to the cooling stage in order to allow characterization of the trap at temperatures up to \SI{300}{\kelvin}. % heating of the thermistor for 
This thermistor permitted measurements of the trap temperature $T_{\rm{trap}}$, calibrated as described in \ref{sec:tempcal}, as well as application of a current to heat the trap independently of the entire base cooling stage.

% to accommodate an available resonator while reducing \ac{RF} dissipation and trap temperature (see Section \ref{sec:expt}).
Application of an \ac{RF} drive frequency $\Omega_{\rm{RF}}/(2 \pi) \sim 20.6$\,MHz at $V_{\rm{rf}} = \SI{183}{\volt}$ introduced a heat load $P \sim 1$\,W in both traps, likely originating from dissipation on the trap.
In trap \#2, where $T_{\rm{trap}}$ was measurable, this raised $T_{\rm{trap}}$ to \SI{185}{\kelvin}.
% With an \ac{RF} drive amplitude $V_{\rm{rf}} = 183$\,V applied to trap \#2, dissipation of $P \sim 1$\,W increased trap temperature to $T_{\rm{trap}} = 185$\,K. % 183 V <- 7.9 dBm -> 189 K; 155 V <- 6.5 dBm -> 155 K
The \ac{RF} dissipation may come from \ac{DC} resistance of the \ac{RF} electrodes (about \SI{1}{\ohm} at \SI{300}{\kelvin}) as well as dielectric loss from the large trap capacitance (about \SI{60}{\pico\farad}), which could be mitigated through adjustments to materials and design geometry.
We observed that our \ac{RF} drive applied across this capacitance also produced large oscillating magnetic fields resulting in \acs{AC} Zeeman shifts \cite{Gan2018} of the levels in the $4 s^2 \rm{S}_{1/2} \leftrightarrow 3 d^2 \rm{D}_{5/2}$ transitions that we used for qubit manipulation.
These shifts varied as the \ac{RF} drive power was adjusted, giving rise to a 3.5\% correction factor for a transition with $\Delta m_s=2$ (denoting spin quantum number $m_s$) for our values of \ac{RF} drive and magnetic field.
Our calculated magnitude of \acs{AC} magnetic field at $V_{\rm{rf}} = 175$\,V, $|\textbf{B}_{\rm{rf}}|=2.96$\,G, is about five times larger than has been reported in another microfabricated trap where this effect was also observed \cite{Zhang2021}. % much larger 
% Chi reported -10 kHz Zeeman shift for S1/2, we get -342 kHz.
% We see 41 kHz/G, -63 kHz/G, 30 kHz/G shifts relative to +1/2->+3/2, -1/2->+5/2, and +1/2->+5/2 transitions (where G is static B-field), which is 2.8%, 1.0%, and 3.5% shifts relative to normal splittings
% We traced this back to our large trap capacitance (about 60\,pF) across which our \ac{RF} drive produced large oscillating magnetic fields resulting in \acs{AC} Zeeman shifts of the relevant levels.
Most measurements required that we properly account for this effect. % , we could collect data dependent on only the parameter of interest. % PS: how related to HR? -- related to everything with 729, actually. especially when changing RF amp. Also couples in B-field noise.

Though we did not perform a systematic study, we observed that single ion storage times were shorter at higher trap temperatures, ranging from days (at trap temperatures close to $T_{\rm{trap}} = 185$\,K) to hours (near $T_{\rm{trap}} = 300$\,K).
However, we did note that at a trap temperature $T_{\rm{trap}} = 83$\,K, strings of more than 10~ions could be trapped stably for days without ion loss. % here, we cooled much redder to allow large excursions to occur without ion loss, but of course this was not optimal for measurements
These three temperatures, from lowest to highest, corresponded with steady-state \ac{OVC} of $1.4$, $1.6$, and $2.6 \times 10^{-8}$\,mbar as measured on a pressure gauge.
This suggests that particles may have been generated near the area of increased temperature --- perhaps desorbed from trap or heater surfaces --- increasing the pressure locally, then eventually registering on the chamber pressure gauge.
Ion loss may relate to collisions stemming from more plentiful or energetic particles \cite{Pagano2018}.
These general observations occurred when either 0.2\,eV- or 1.0\,eV-depth voltage sets were applied, and so some effect besides trap depth, such as laser cooling capability, was more likely the primary factor contributing to ion loss.

\subsection{Probing motional mode frequencies} % Demonstration of 1\,eV trap depth
\label{sec:depth}

To identify secular motional frequencies of a single trapped ion, we apply a voltage configuration, compensate micromotion, and then perform spectroscopy on the motional sidebands of a chosen $4 s^2 \rm{S}_{1/2} \leftrightarrow 3 d^2 \rm{D}_{5/2}$ transition.
Here we present the data for trap \#2, for which the most extensive characterization was performed.
From the 1\,eV voltage set in Table~\ref{tab:voltages}, we simulate mode frequencies $(0.88, 3.80, 3.76)$\,MHz, ordered by the mode vectors closest to $(\hat{x}, \hat{y}, \hat{z})$.
Applying this voltage set, measurement of the trap gives frequencies $(0.81, 3.74, 3.80)$\,MHz, consistent with the expected potential to within 8\% axially and 2\% radially (Figure~\ref{fig:matching}b, `X' markers).
We next applied the 0.2\,eV voltage set in Table~\ref{tab:voltages}, whose lower voltages provide more leeway for parameter variation without reaching voltage limits, and varied \ac{RF} amplitude (changing radial frequencies) and \ac{DC} voltages that modify axial curvature (changing axial frequencies). % We also vary \ac{DC} and \ac{RF} voltages over a larger parameter space using voltages close .
Over a set of parameters producing frequencies $f_{\rm{meas}}$ in the range 0.6--3.8\,MHz, measured frequencies matched with simulated values $f_{\rm{sim}}$ to within 10\% for the axial mode and 5\% for radial modes (Figure~\ref{fig:matching}b).

Uncertainty in applied $V_{\rm{rf}}$ likely dominates the radial frequency mismatch in Figure~\ref{fig:matching}b.
Stray electric field curvature, which is not included in these simulations, would explain a deviation in axial frequencies.
Using sideband spectroscopy, we measured a residual anti-confinement (corresponding with axial frequency $\omega / 2 \pi = -45(3)\,\rm{kHz}$) at the center trapping site in trap \#2. % that agrees with the axial field curvature.
Possible sources of such fields are discussed in Section \ref{sec:field}. % stray field particularly manifests in an axial frequency difference
% Our observations, consistent with large trap depth, show that microfabricated \ac{3D} trap fabrication is an effective method by which to scale ion trap complexity.
% Our observations, consistent with large trap depth, confirm that the chosen microfabricated 3D trap fabrication is an effective method to produce high-depth traps. % model of a deep, stable 3D trap.
In any case, these observations validate the designed model over a wide range of parameters, consistent with high-depth traps that meet the design goals of our \ac{3D} microfabrication process. % the chosen method of 3D microfabrication is effective to produce high-depth traps.

\subsection{Measuring stray static electric fields}
\label{sec:field}

\begin{figure}[htb]
    \centering
    \includegraphics[width=1\textwidth]{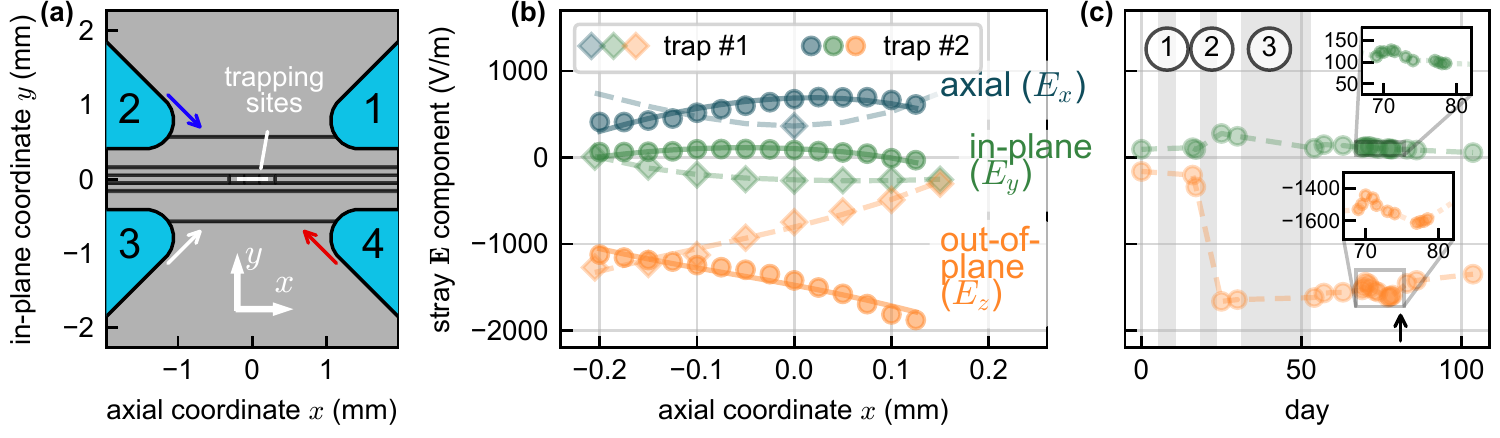}
    \caption{Measurement of stray static electric field components $E_x$, $E_y$, and $E_z$ along the trap axis $x$ and over time. 
    a) Bottom wafer geometry (gray) overlaid with the glass spacers (blue) labeled 1--4.
    In the text, we assess how well data match with models of various stray field sources, including the side facets of these spacers and dielectric between or on top of the electrodes, to explain measured fields.
    Arrows depict the paths of the high-power \SI{729}{\nano\meter} beam (red), remaining beams (blue), and neutral flux (white).
    b) Stray field is measured for the three field components along a range of trapping positions by using micromotion compensation techniques to null positional offsets.
    Fit results (trap \#1: dashed; trap \#2: solid) to simulated field produced by each spacer can account for field curvature, provided large static offset fields are present (see text).
    c) Radial stray field components at $x=0$ are monitored in trap \#2 over months of continuous trap operation.
    Special events (gray regions) interrupted routine data-taking: (1) thermally cycling from base temperature to 300 K and back, while vacuum pumping, (2) adjusting and reconnecting \ac{DC} circuitry, and (3) suffering an extended electrical outage that degraded vacuum, warmed and cooled the cryostat, and reset electrical systems, and (insets) repeated cycling of trap temperature between days 65--80.
    The arrow marks the time when data was collected for (b).
    Dotted lines connect data points to serve as a guide to the eye.
    }
    \label{fig:field}
\end{figure}

Dielectric surfaces that are not electrically shielded from the ion can generate stray electric fields \cite{Teller2021}.
Stray fields that are not sufficiently stable or homogeneous may not be compensated well enough to perform high-fidelity quantum gate operations \cite{BallanceThesis}, and could make it difficult to scale this technology.
% Relative to its single-wafer predecessor using the same fabrication technology \cite{Holz2020}, the \ac{3D} trap reported here introduces additional dielectric layers, in the form of the spacer and top wafer dielectrics, as described in Section \ref{sec:fabrication}.
% However, this design generally shields bulk dielectric using ground planes and undercuts also adopted in other microfabricated traps \cite{Britton2009,Maunz2015,Jung2021}.
% In any case, the need to pattern electrodes, and thus form gaps between them, means that some dielectric surfaces remain exposed to the ion in close-packed geometries. % could condense this paragraph if necessary
To measure stray fields, we primarily use parametric-excitation-based micromotion compensation techniques \cite{Wineland1998}.
From trap simulations, we infer the stray field corresponding to the applied micromotion compensation voltages.
Our implementation is sensitive to $\sim 3$\,V/m radial and $< 1$\,V/m axial static electric fields. % (though we are implementing time-tagging techniques \cite{Nadlinger2021} to improve this further).
We characterize the fields present in both traps along the axis at different trapping sites (Figure~\ref{fig:field}a).

% Understanding sources of stray field can also reveal sources of electric field noise that affect ion heating rates.

% proposed order:
% we see this data
% data is similar to what's seen elsewhere
% the following sources are generally used to explain these observations, and can be used here as well
% here are where the particular sources may apply to our geometry, and what's special
% then,
% changes over time

% here: general characteristics of data, particularly the offsets
The results (Figure~\ref{fig:field}b) show large offset field components (up to about $1.5 \times 10^3$\,V/m in the out-of-plane direction) with about $1 \times 10^3$\,V/m axial variation along the measured distance range of $\pm 0.2$\,mm from the trap center.
% The strong axial field component near the trap center and moderate axial curvature (in trap \#2, a residual anti-confinement corresponding with axial frequency $\omega / 2 \pi = -45(3)\,\rm{kHz}$) suggests a combination of local and distributed field sources. % I wouldn't say this is what clinches it
The magnitude of field components in both traps is similar to many single-wafer surface traps \cite{Allcock2010,Merrill2011,Narayanan2011,Doret2012,Clark2014,Shu2014,Mehta2020,Holz2020} made from a variety of materials and measured at different temperatures and ion-surface distances. % discussing similarities
Proposed sources of such fields include charge present on dielectric or metal surfaces \cite{Harlander2010,Wang2011,Ong2020}, as well as patch potentials comprising surface oxides \cite{Allcock2011,Teller2021} or adsorbates \cite{McGuirk2004,Obrecht2007,Brama2012}. % now discussing sources
% Here we discuss the most probable sources, in particular to determine whether they may be related to the trap design.
These sources can contribute to field offsets if they are distant or distributed, or to field curvature if they are close and point-like.
% discuss why Z is much bigger than X and Y?

Since the stray field is inferred from the micromotion compensation required to move the ion to the \ac{RF} null, trap misalignment that offsets the \ac{RF} null can be mistaken for an offset field.
This primarily affects lateral ($\hat{y}$) offsets, which from the measured misalignment gives a standard deviation $\Delta E_y = 59$\,V/m.
Uncertainty in the position of the null above the surface is dominated by spacer wafer thickness, $\SI{400(3)}{\micro\meter}$, to give $\Delta E_z = 48$\,V/m.
The measured misalignment contributes negligibly to axial field uncertainty. % because of the method used, and the small effect of misalignment in that direction
While misalignment could account for part of the offset of the measured fields, it cannot explain deviations larger than $\Delta E_y$ or $\Delta E_z$ over the range of trapping positions.

% The size and distance of dielectric and conducting surfaces in our trap is similar to many surface-electrode traps.
% Relative to these, however, the presence of top and space wafers introduces concerns that fields may be higher.
% However, we observed fields that were commensurate with many other traps.
% We used a model of spacer sidewall charge to fit the measured data; however, other sources of charge would likely produce similar effects.
% here, explain the fits and the meaning of variation of the data
In contrast to surface-electrode traps, one possibly significant source of electric field could be charges on the sidewalls of the spacer wafer.
The data in Figure~\ref{fig:field}b (lines) are fitted to the simulated field from uniform charge densities on these sidewalls, producing values up to 2500 elementary charges per $\SI{}{\micro\meter}^2$ (Table~\ref{tab:field_fits}).
% The fitted data produces the values and uncertainties given in Table~\ref{tab:field_fits}. % $(-2120(340), 860(400), 1840(380), 390(260)) e^{-}/\SI{}{\micro\meter}^2$ on spacers 1--4 for trap \#1 and $(1380(200), 630(360), -2510(450), 320(250)) e^{-}/\SI{}{\micro\meter}^2$ on spacers 1--4 for trap \#2, and $\mathbf{E}_{\rm{offset}} = (-450(50), -380(50), -900(170))$ V/m in trap \#1 and $(1310(40), 210(50), -1500(40))$ V/m in trap \#2.
This is larger than the charge densities reported on fiber surface near ions \cite{Ong2020} and on a planar trap \cite{Harlander2010}, but these materials and geometries may not be comparable.
The model uses seven free parameters: a uniform charge density on each of the spacers 1--4 that control field curvature along the axis, and the three components of a constant, arbitrary field $\mathbf{E}_{\rm{offset}}$ that set the offset.
% Differences between the two traps could come from local variations in charge distribution. % paragraph below, where to put it?
Additional variability of 400\,$e^{-}/\SI{}{\micro\meter}^2$ in charge density and 200\,V/m in $|\mathbf{E}_{\rm{offset}}|$ may come from simulated field uncertainty, considering that the spiked sidewall's position (Figure~\ref{fig:fabrication}d) has a $\SI{10}{\micro\meter}$ standard deviation due to an uncertain etch profile.

The field offsets could be explained, for example, by a charge density of $10^5\, \e^{-}/\SI{}{\micro\meter}^2$ in the gaps between electrodes on the bottom wafer, which we find in simulation to produce an out-of-plane field offset close to the $E_{\rm{offset},z} \sim -1000$\,V/m measured at the center trapping site.
Though this exceeds the charge density that would be required on the spacer sidewalls to produce a comparable field, this field magnitude is nonetheless common among microfabricated surface-electrode traps without significant exposed bulk dielectric \cite{Allcock2010,Mehta2020,Holz2020}, and so such surface-based field sources cannot be ruled out.
Due to the relative asymmetry of the trap in the out-of-plane direction, these effects all tend to produce larger field components along $\hat{z}$, which is consistent with the measured data. %, could break the symmetry, explaining the smaller offsets in the the $\hat{x}$ and $\hat{y}$ directions that accompany the large out-of-plane ($\hat{z}$) field. %
If this charge were unevenly distributed, or debris were present on a trap surface, this could provide curvature in addition to an offset field.

Offset fields might also come from far-away structures that are not part of the trap assembly, like charged lenses or windows. %, which could explain large, axially invariant offset fields.
Calculations predict that fields from the nearest dielectric surface (43\,mm away) would be shielded by a factor of $\sim 8$ by the largest aperture ($\sim 3$\,mm) between the dielectric surface and the ion.
The trap is otherwise surrounded by grounded conductors.
Therefore, sources of electric field \emph{within} the shielded region are most consistent with the observations.

\begin{table}[htb]
\centering
\caption{
    The results of fitting axial stray field data in both measured traps to a model of uniform charge density on each of the spacers 1--4 and an arbitrary field, $\mathbf{E}_{\rm{offset}}$.
    Values in parentheses represent the standard error from the fit.
    }
\begin{tabular}{lrr}
\toprule
{} &  trap \#1 &  trap \#2 \\
\midrule
charge densities $(e^{-}/\SI{}{\micro\meter}^2)$ & & \\
\midrule
spacer 1            &          -2120(340)   &           1380(200) \\
spacer 2            &          860(400)     &           630(360) \\
spacer 3            &          1840(380)    &           -2510(450) \\
spacer 4            &          390(260)     &           320(250) \\
\midrule
offset fields (V/m) & & \\
\midrule
$E_{\rm{offset},x}$   &         -450(50)      &           1310(40) \\
$E_{\rm{offset},y}$   &         -380(50)      &           210(50) \\
$E_{\rm{offset},z}$   &         -900(170)     &           -1500(40) \\
\bottomrule
\end{tabular}
\label{tab:field_fits}
\end{table}

% We see a large offset, and we see curvature
% Offset could come from uniform gap or surface charge, same as other traps
% Curvature could come from junk, sidewalls, or also gaps if distribution is local

% sources of constant offset

% For example, we separately calculate that a charge density of $10^5\, \e^{-}/\SI{}{\micro\meter}^2$ in the gaps between electrodes on the bottom wafer would be required to produce an out-of-plane field offset close to the $E_{\rm{offset},z} \sim -1000$\,V/m measured at the center trapping site.

% Uneven charging of gap dielectric surfaces, or debris on the surface away from the center of the trap, could break the symmetry, explaining the smaller offsets in the the $\hat{x}$ and $\hat{y}$ directions that accompany the large out-of-plane ($\hat{z}$) field. % explains why there is z asymmetry

% Using sideband spectroscopy, we measure a residual anti-confinement (corresponding with axial frequency $\omega / 2 \pi = -45(3)\,\rm{kHz}$) at the center trapping site in trap \#2 that agrees with the axial field curvature.
% for the record, trap #1 had either +57(6) kHz or -8(3) kHz curvature depending on the range we examined... there was some kind of a discontinuity so I don't know what to trust and don't include it. If it were in fact +57 then that would go along nicely with the axial curvature, but no way to know.

The sources responsible for these electric fields may be intrinsic, induced by light, or introduced by atomic flux while operating the trap.
Some light from the laser beams that pass through the trap (Figure~\ref{fig:field}a, blue and red arrows) was observed to scatter from the trap's inner surfaces and the sides of spacers 2 and 4.
If this induced charge, we would expect the strongest effect from the deepest-\ac{UV} wavelength, \SI{375}{\nano\meter} photoionization light, present at \SI{30}{\micro\watt} during loading.
During loading, neutral calcium also travels along the direction indicated by the white arrow in Figure~\ref{fig:field}a, and can deposit on trap surfaces including the sides of spacers 1 and 3.
Adsorbate build-up has been shown to alter stray fields \cite{Zhang2020a,Zhang2021}.
However, we observe that the field remains stable during trap operation, and we do not observe significant stray field drift in the seconds or hours after loading (resolvable by tracking the ion’s position on a camera). % and is largely independent of trap temperature. -- will discuss this later % despite exposure to \ac{UV} beams and atomic flux, 
Therefore, while induced charge on these surfaces may contribute to part of the measured field, the bulk of the charge appears to be intrinsic.

We observed small day-to-day changes, culminating in weekly field drifts of 10\,V/m in-plane and 100\,V/m out-of-plane (Figure~\ref{fig:field}c), which could be explained by local charging and dissipation or fluctuations of \ac{DC} voltages.
These fluctuations were small relative to the axial position field dependence in Figure~\ref{fig:field}b.
Special events like long-term electrical outages, or vacuum or thermal cycling, produced changes at a faster rate.
In one such case, a significant change in measured field occurred after \ac{DC} electronics were disconnected and reconnected.
This could have possibly reconnected floating electrodes (thus changing the calibration of applied compensation or affecting the charge state of weakly implanted electrons in dielectric close to the electrodes \cite{Ong2020}) or changed the contact potential at a conductor interface.
We note that after the jump the compensated field drift was observed to reverse direction. %  and trend towards its former value.
None of these events are likely to have removed adsorbed surface contaminants.
% Rather, it seems plausible that our observed weekly field drifts of 10\,V/m in-plane and 100\,V/m out-of-plane fluctuations arose from changing surface charge distributions. 
% However, we do not see systematic drift over long time periods (Figure~\ref{fig:field}c) that would indicate adsorbate build-up. % to have a dominant effect.

Overall, the behavior of stray electric fields is similar to that seen in other microfabricated traps, which is consistent with models of charge on bulk dielectric (in our case, for example, spacer sidewalls), in dielectric gaps, or on conductor surfaces in the electrode plane. % and with broadly distributed, mostly static surface charge distributions.
% A large portion can be attributed to charges on the surface of dielectric in the gaps between electrodes and on the surfaces of spacer sidewalls.
The observed fields exhibit low drift and are well within the range that can be compensated using low voltages ($\ll 10$\,V) applied to the available \ac{DC} electrodes, and thus they do not present an obstacle to trap operation that could hinder the scaling prospects of our \ac{3D} \ac{MEMS} ion trap technology.

\subsection{Heating rates}
\label{sec:hr}

Motional heating of ions reduces the fidelity of single-qubit gates and multi-qubit entangling operations \cite{BallanceThesis} that are critical to a trapped ion quantum processor \cite{Monroe1995}.
Heating of the ions' motional modes occurs at a rate $\Gamma_h$ due to electric field noise \cite{Turchette2000}.
This rate can be related to the electric-field noise spectral density, $S_{E}$, using \cite{Brownnutt2015}
\begin{equation}
    \label{eq:specdensity}
    S_{E} (\omega_m) = \frac{4 m \hbar \omega_m}{e^2} \Gamma_h (\omega_m)
\end{equation}
where $m$ is the ion mass, $\omega_m$ the mode frequency, and $e$ the electron charge.
We measure heating rates of a single ion’s axial, in-plane radial, and out-of-plane radial motional modes using the sideband ratio method \cite{Leibfried2003}. % ratio of sideband Rabi excitations
By comparing the $\Gamma_h$ (or $S_{E}$) dependence on mode frequency, trap temperature, and system configuration to known models, we can pinpoint noise sources and evaluate trap performance. % , , relative to the trap design , and place bounds on the fidelity of operations that may be performed. % we don't actively state a fidelity anywhere, this might be an issue
% We vary mode frequency, mode rotation angle, trap temperature, and electrical configuration. % this sentence is now redundant because we have "system configuration" above
In addition, we monitor drift and scatter of heating rates over time, as well as the response to micromotion miscalibration. % and verify that measurements do not change significantly when we apply radial electric fields up to $\sim 75$\,V/m that displace the ion about $\SI{0.5}{\micro\meter}$ from the \ac{RF} null, which examines whether \ac{RF} noise contributes to the heating rate. %, which corresponds with many times the uncertainty of our micromotion compensation.

\begin{figure}[htb]
    \centering
    \includegraphics[width=1\textwidth]{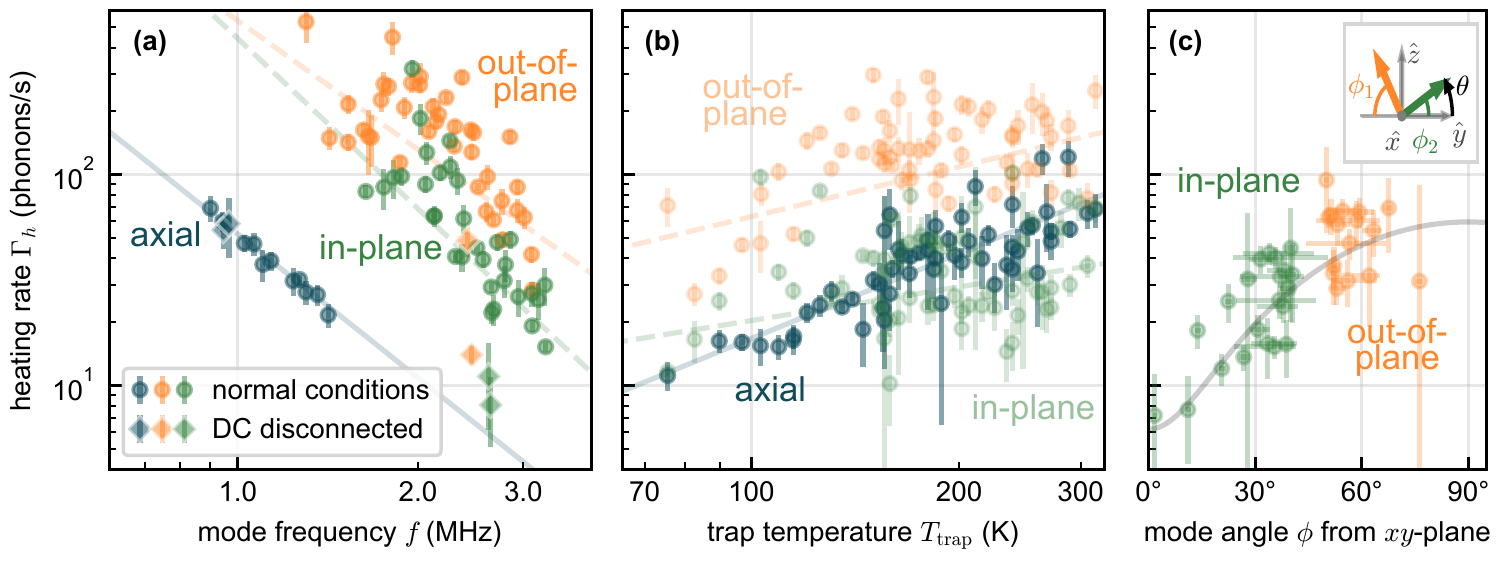}
    \caption{Heating rates are measured for axial (blue circles) and radial modes (green, orange circles) as a function of mode frequency, radial mode rotation, and trap temperature.
    a) Axially confining \ac{DC} voltages are varied to adjust axial mode frequency, while the \ac{RF} power is varied to adjust radial mode frequencies, with mode rotation angle $\theta = 38^{\circ}$.
    % Mode values are normalized (see text) to a trap temperature of 185\,K using the temperature dependence in (b).
    When \ac{DC} voltage sources are disconnected, lower radial rates (diamonds) are measured.
    In this case radial heating rates align more closely with the axial power-law scaling (solid blue line, exponent $-2.3(1)$), consistent with limitation from surface noise sources (see text).
    Fits to power law models (dashed) give exponents $-2.7(3)$ and $-2.0(3)$ for in-plane and out-of-plane radial modes, respectively.
    b) Trap temperature is varied using a heater near the base of the trap chip.
    Values are measured at $\theta = 38^{\circ}$ and normalized (see text) using the frequency scalings in (a) to data set mean frequencies $(0.98, 2.35, 2.56)$\,MHz, for axial, out-of-plane, and in-plane modes respectively.
    Temperature scalings are fit to a power-law model (lines) give exponents $(1.34(8), 0.5(2), 0.8(1))$.
    Radial mode data are only weakly correlated with an exponential temperature scaling model (dashed lines), consistent with a limitation from noise sources that are external to the trap.
    The opacity of radial data is reduced to improve clarity.
    c) \ac{DC} voltages adjustment rotates radial modes by $\theta$ (see inset), giving each mode an angle $\phi$ with respect to the $xy$-plane.
    The resulting measured heating rates are fit (line) to a sinusoidal dependence. %, suggesting a strong out-of-plane heating component. % from possible external noise sources.
    Measurements are taken at $T_{\rm{trap}} = 185$\,K, and frequencies are normalized (see text) to data set mean frequency 3.0\,MHz. % using the axial mode frequency scaling in (a). 
    Some outliers are omitted for clarity (see text).}
    \label{fig:hr}
\end{figure}

Measured heating rate data are shown in Figure~\ref{fig:hr}, plotted against mode frequency $f_m = \omega_m / 2 \pi$, trap temperature $T_{\rm{trap}}$, and radial mode angle $\phi$.
To access all modes with laser beams propagating parallel to the plane of the trap, we employ voltage sets (starting with the 0.2\,eV-deep set in Table~\ref{tab:voltages}) that rotate the radial modes by an angle $\theta = 38^{\circ}$ relative to $\hat{y}$ and $\hat{z}$ using the off-axis shim electrodes. % access might be vague. also maybe "and cool"
%, show that this trap's heating rates are improved relative to its predecessor \cite{Holz2020}, perhaps due to improved ground plane coverage and dielectric shielding.
The lowest measured heating rates are below 10\,phonons/s at 2.6\,MHz at $T_{\rm{trap}} = 165$\,K and around 40\,phonons/s at 1.0\,MHz at $T_{\rm{trap}} = 185$\,K; using Eq. \ref{eq:specdensity} these evaluate to electric field spectral noise densities of $S_{E} = 1.8 \times 10^{-13}$ and $2.8 \times 10^{-13}$ $\rm{V}^{2} \rm{m}^{-2} \rm{Hz}^{-1}$, respectively. % f-scaling dominates over T-scaling, so higher f generally better, but our best measurements were in the noise-suppressed regime, which we did at only one T and f... I think it's obvious from the plot
% Kirill had S_E = 5.2e-16 for SC trap, https://journals.aps.org/pra/abstract/10.1103/PhysRevA.99.023405
% The trap has an ion-electrode distance of $\SI{200}{\micro\meter}$, and data is taken at a trap temperature 185\,K. % 185 is too general
These heating rates are close to those measured elsewhere \cite{Brownnutt2015, Brown2021} when comparing to traps with similar ion-electrode distance ($\SI{200}{\micro\meter}$).
These measurements were taken in trap \#2.
While trap \#1 produced heating rates as low as 26(2)\,phonons/s for a radial mode at 2.6\,MHz, the trap temperature could not be estimated on this device, and calibrations were not performed as systematically as for trap \#2.
Therefore, all heating rate data presented here was measured using trap \#2.
% Adjustment of the trap \ac{RF} was the primary mechanism by which we changed trap temperature and radial mode frequency, ...
% Rates are low despite the addition of dielectrics and the relatively large stray electric fields described in Section \ref{sec:field}. % here I mean the spacers and extra material from more wafers... maybe this won't get through

In Figure~\ref{fig:hr}a--b, axial mode data exhibit low scatter, while radial mode data are more scattered relative to the low uncertainty at each point.
The uncertainty of individual measurements is obtained by applying binomial statistics to a series of `bright' or `dark' determinations from thresholding photon detection events within a detection window \cite{Negnevitsky2018}.
The plotted uncertainties represent the standard errors of the mean of fitted heating rates for multiple such measurements.
The axial mode heating rate follows a power-law scaling model with respect to both temperature and frequency, $\Gamma_{h} \propto f^{-\alpha} T_{\rm{trap}}^{\beta}$ for exponents of mode frequency ($f$, Figure~\ref{fig:hr}a) and temperature ($T_{\rm{trap}}$, Figure~\ref{fig:hr}b).
Fits of the axial heating rate data to $A f^{-\alpha}$ and $B T_{\rm{trap}}^{\beta}$ give $\alpha = 2.3(1)$ and $\beta = 1.34(8)$ respectively.
Literature values are often given in terms of $S_{E}$ scaling coefficients $\alpha'$ and $\beta'$, which are related through $S_{E} \propto \omega_m \Gamma_{h} \propto f^{-(\alpha-1)} T_{\rm{trap}}^{\beta}$ to give $\alpha' = \alpha - 1 = 1.3(1)$ and $\beta' = \beta = 1.34(8)$.
% Regardless, lower temperatures (such as through better thermal anchoring) would provide an overall benefit that we do not demonstrate here.
Our scaling exponents are similar to those given in other work \cite{Kim2010,Wang2010a,Allcock2011,Teller2021}, in particular the $\alpha' \sim 1$, $\beta' \sim 1$ seen to result from electric field noise from surfaces, for example as in ref. \cite{Kumph2016}. % some of these initial refs aren't fully applicable here
% Our results are close to $\alpha' \sim 1$, $\beta' \sim 1$ seen to result from electric field noise from dielectric surfaces \cite{Kumph2016}.
% They follow reported observations from many other surface traps and the scaling exponents are consistent with proposed models of anomalous heating \cite{Kim2010,Wang2010a,Allcock2011,Teller2021}.
We cannot access low enough trap temperatures to distinguish whether the temperature scaling $\beta$ remains unchanged below our lowest measured temperature, 75\,K, or the temperature scaling is disjointed as in ref. \cite{Labaziewicz2008,Chiaverini2014}.

% Our axial heating rates also follow a power-law temperature dependence (Figure~\ref{fig:hr}b). % this is obvious from the alpha and beta given above
The scatter in radial heating rates in Figure~\ref{fig:hr}a is consistent with the effect of a ``technical'' noise source, with many distinct spectral features that collectively contribute to a large background rate. % we also see this in coherence measurements
In the presence of such a source, one would expect an ion to exhibit higher heating rates when its motional frequency is tuned closer to frequency peaks in the electric field noise spectrum.
% Temperature variation also leads to small frequency shifts of the \ac{RF} resonator circuit. % this is a bit confusing still, why does this change the susceptibility of modes to the noise spectrum?
When fit to a power law model, we find exponents $-2.7(3)$ and $-2.0(3)$ for in-plane and out-of-plane modes; the expected frequency dependence would depend on the nature of the external noise source.
This scatter behavior is also consistent with data in Figure~\ref{fig:hr}b, where radial heating rates are not significantly correlated with trap temperature that was adjusted using a heater.
In this case, one would expect trap temperature to affect modes for which surface-based noise sources limit the heating rate, while trap temperature would hardly change the effect of external sources. % except for the RF reflection characteristics of the trap, of course, which we know changes the mode frequencies and the amount of power/heat delivered, etc...
This points to an externally generated noise source, which may be carried along transmission lines to the \ac{DC} electrodes \cite{Brownnutt2015}.
% The radial heating rates in Figure~\ref{fig:hr}b are not significantly correlated with temperature, a behavior consistent with noise limited by external sources.

Normalization of heating rates was performed in some cases to easily compare data sets collected where different \ac{DC} or \ac{RF} voltages were applied, or where trap impedance was modified by the temperature, thus slightly changing delivered power, altering \ac{RF} amplitude, and shifting motional mode frequencies (see \ref{sec:normalization}).
Heating rates measured while sweeping the trap temperature in Figure~\ref{fig:hr}b were frequency-normalized to the mean frequency values of axial, out-of-plane radial, and in-plane radial modes, $(0.98, 2.35, 2.56)$\,MHz.
Heating rates in Figure \ref{fig:hr}a were not temperature-normalized, however, since axial mode rates were taken at a constant temperature and radial mode rates were not strongly correlated with temperature (\ref{sec:normalization}).
% Axial heating rates in Figure~\ref{fig:hr}a were measured at a constant $T_{\rm{trap}} = 185$\,K, and did not require normalization.
% However, we adjusted the \ac{RF} voltage in Figure~\ref{fig:hr}a to change radial mode frequencies, which caused the trap temperature to vary between 75\,K and 193\,K.

% Normalization is important for comparing data that depends on many parameters, but we find that without it, the qualitative behavior of the data in Figure \ref{fig:hr} is retained.

We noted that the scatter of radial mode heating rates in Figure~\ref{fig:hr}a--b was consistent with a technical noise spectrum that may originate externally; however, temporal variation of noise sources could also produce scatter.
To check this, we repeatedly interleaved calibrations and heating rate measurements over nearly 8~hours (with 8~measurements at 43~minute intervals followed by 10~measurements at 13~minute intervals) at $T_{\rm{trap}} = 153$\,K and at mode frequencies $(0.95, 2.46, 2.65)$ MHz.
Over 19 sequential measurements, this gave mean values of $41$, $106$, and $20$\,phonons/s for axial, out-of-plane radial, and in-plane radial modes, and standard deviations $22$, $9$, and $7$\,phonons/s, respectively (see \ref{sec:uncertain_hr}). % data in https://gitlab.phys.ethz.ch/tiqi-projects/waveguides/notebooks/-/blob/master/PIEDMONS_paper/HR_configs.ipynb
Despite significant axial rate uncertainty under these conditions, the standard deviation of radial rates is small.
This suggests that radial heating rate scatter is not the indirect product of drifts in noise properties over time.
Seven measurements were performed in the presence of radial shimming fields up to $\pm 45$\,V/m, which displaced the ion about $\SI{0.5}{\micro\meter}$ from the \ac{RF} null.
They did not show a strong dependence on the stray field magnitude, and gave respective mean values $44$, $108$, and $28$\,phonons/s, and standard deviations $17$, $6$, and $14$\,phonons/s, respectively. % data in https://gitlab.phys.ethz.ch/tiqi-projects/waveguides/notebooks/-/blob/master/PIEDMONS_paper/HR_configs.ipynb
This suggests that neither noise delivered through the \ac{RF} electrodes nor drifts in static stray fields over time play a significant role in the measured heating rates.
The data in Figure~\ref{fig:hr}a--b are taken at an angle $\theta = 38^{\circ}$ of the radial mode vectors about the trap axis.
To understand the effect of this angle, we applied \ac{DC} voltage sets that adjusted $\theta$ from $0^{\circ}$ to $45^{\circ}$.
This sets each mode's angle $\phi$ with respect to the $xy$-plane (Figure~\ref{fig:hr}c inset).
The ion excitation signal while driving motional sideband transitions also depends on this angle through an effective Rabi rate sensitive to the angle between the beam wavevector and the mode \cite{Kienzler2015}.
From these measurements, we cross-checked the designed angle $\phi$ and obtained uncertainty values.
% To check the resulting angle, and to obtain an uncertainty value, we determine the angle between the beam wavevector and the mode by driving motional sideband transitions and fitting the excitation signal in time.
% This signal depends on an angle-sensitive effective Rabi rate \cite{Kienzler2015}. % result at this point: about +/- 0.1 in tilt voltage factor (thus variable angle uncertainty with tanh dependence)
As a mode approaches $\phi = 90^{\circ}$ (out of the plane and orthogonal to the cooling beams) sideband cooling performance degrades, higher initial photon numbers are measured, and the uncertainty of heating rates using the sideband ratio method increases.
Therefore, for each mode, data collection is conditioned on cooling to initial values of average photon number $\bar{n} \leq 1.5$, and also on confidence in mode frequency from a calibration using sideband spectroscopy (requiring standard error $< \SI{100}{\kilo\hertz}$).
Even with these conditions, a few outliers result (for example, with negative rates or \textgreater 400\% relative uncertainty), which are omitted from the plots for clarity.
To directly examine the dependence of heating rate on $\phi$, frequencies are normalized to the mean frequency of both radial modes' data (3.00\,MHz) using the procedure described above. %!!!
% The radial modes' frequencies have mean values $(2.91, 3.05)$\,MHz and standard deviations $(0.03, 0.02)$\,MHz, leading to average normalization factors $(0.90, 1.08)$.
The resulting data (Figure~\ref{fig:hr}c) show that heating rates increase significantly as $\phi$ rotates from an in-plane orientation (along $\hat{y}$) to an out-of-plane orientation (along $\hat{z}$).

These measured heating rates can be described by a sinusoidal model that gives rates at $\phi=0^{\circ}$ ten times lower than at $\phi=90^{\circ}$.
If electric field noise at the ion were limited by losses in a surface dielectric or metal film, one would expect out-of-plane rates two times larger than in-plane rates \cite{Kumph2016}.
% This strong surface polarization, along with the scattered radial heating rates in Figure~\ref{fig:hr}a--b and stable radial rates over time, are consistent with a ``technical'' noise source that has many distinct spectral features that collectively contribute to a heightened background rate. % we also see this in coherence measurements
% In the presence of such a source, one would expect the ion to exhibit higher heating rates as its motional frequency is tuned closer to frequency peaks in the electric field noise spectrum. % cite?
% Surface polarization behavior could represent electric field noise oriented mostly along $\hat{z}$, which could stem from voltage noise on the electrodes on the top and bottom wafers
The observed excess scaling, however, is more consistent with a model of noise delivered from an external source, such as through trap electrodes.
By symmetry of this trap's electrode layout and its asymmetry in the out-of-plane direction, voltage noise common to all \ac{DC} electrodes would produce electric field components that are normal to the electrode planes.
% This is consistent with the observed surface polarization behavior, as well as the scattered behavior described above. % relative field is array([-0.05522702, -0.70347133,  1.        ]) for 1 V on all electrodes, doesn't consider ground

% having analyzed all the collected data, swing over to remediation
Considering whether externally generated noise may couple to the electrodes through \ac{DC} wiring and limit the heating rates measured on radial modes, we adjusted the \ac{DC} electrical configuration (further details given in \ref{sec:elecconfig}).
We found that radial heating rates were only significantly reduced when disconnecting the trap from its \ac{DC} voltage sources \emph{and} ground references (which also reference the \ac{RF} ground).
Disconnection left the electrodes electrically floating, with the capacitance of the electrodes and \ac{DC} filter board capacitors maintaining a nearly constant voltage over time.

With \ac{DC} lines disconnected (Figure~\ref{fig:hr}a, diamond points) the radial heating rates drop, nearly reaching the frequency scaling trend line set by the axial mode.
Even when normalizing for frequency, the average in-plane rate remains below the out-of-plane rate by a factor of 2.3, which is almost consistent with noise due to a bulk dielectric or surface layer noise models \cite{Kumph2016,Teller2021}.
However, the two out-of-plane data points (which are taken at slightly different frequencies and temperatures) still demonstrate scatter exceeding the measurement uncertainties. % 
% This behavior is consistent, however, ... expected from noise due to a dielectric surface layer \cite{Kumph2016}.
This scatter and the discrepancy in the ratio of the radial modes' rates suggest that some technical noise sources may persist and become dominant when the \ac{DC} voltage sources are externally disconnected.
These may include inductive pick-up within the cryostat or noise on the \ac{RF} electrodes.
% Heating rates were measured before, after, and upon reversal of changes to the electrical configuration -- including and only by disconnecting all connections to the \ac{DC} signal sources was the difference in heating rates statistically significant.
% Relative to stronger frequency or ion-surface distance dependencies \cite{Sedlacek2018a,Turchette2000}, trap temperature has been shown to produce a relatively weak heating rate power-law dependence, but can still vary by orders of magnitude over the range from 4\,K to 300\,K \cite{Labaziewicz2008, Chiaverini2014

% some conclusion...
Heating rates of the axial mode were unaffected by the disconnection of an external noise source, which along with its frequency and temperature dependence suggests a limiting noise source originating on trap surfaces or in bulk dielectric.
The relatively low spectral noise densities of this source suggest that our \ac{MEMS} ion trap technology does not face an exceptional obstacle to scaling due to surface material composition or other aspects of the fabrication method.
Radial heating rates, on the other hand, appear to be dominantly set by noise sources originating externally to the trap as evidenced by their scatter, trap temperature independence, strong out-of-plane polarization, and response to \ac{DC} disconnection.
Regardless, for the purpose of trap characterization, these measurements place an upper bound on the contributions from trap surface-based noise sources.
Improvements to the apparatus would reduce technical noise and improve the radial modes' sensitivity to remaining sources of noise, benefiting further trap characterization.
% 
% These values are comparable to those in devices with similar characteristics, and are already satisfactory for operating this trap in numerous applications.
% Upcoming improvements to the apparatus should reduce technical noise and improve sensitivity to the remaining sources of noise, benefiting future characterization.
% This strong surface polarization, along with the scattered radial heating rates in Figure~\ref{fig:hr}a--b and stable radial rates over time, are consistent with a ``technical'' noise source
% say something good about our rates as well? nah, leave it for the absolute conclusion, which follows immediately

% Varied temperature modifies \ac{RF} impedance (see text) and changes power and frequencies slightly; thus, 

\section{Conclusion and Outlook}

We have demonstrated the operation of a microfabricated \ac{MEMS} \ac{3D} ion trap with a trap depth of 1\,eV produced in an industrial facility.
% Stable, long-term trapping of many ions exhibits a hallmark of \ac{3D} traps.
Measurements of motional frequencies agreed to within $\pm 5$\% of predicted values over a broad range of \ac{DC} and \ac{RF} voltages, validating our simulation model and suggesting that the trap was accurately fabricated. % supporting the statement of accurate fabrication ... % suggesting that the trap simulations were accurate and that the fabrication process accurately realized a trap with high depth.
We found motional heating rates (\textless40\,phonons/s at 1\,MHz axial frequency) commensurate with electric field noise spectral densities found in surface traps of this scale \cite{Brownnutt2015, Brown2021}. % have to be careful here
Our traps could be operated stably, with strings of more than 10~ions trapped for days at lower trap temperatures. % storage time
Though the data presented was only measured at trap temperatures as low as 70\,K due to \ac{RF} dissipation on the trap, stronger thermal anchoring to the base cooling stage operating with 1.5\,W of cooling power at $T_{\rm{base}} = 6.5$\,K or changes to trap materials and geometry should further reduce heat load. % our measured heating rate trend predicts $< 1$\,phonon/s at 1\,MHz at a base temperature $\text{T} \sim$ 10\,K.
We also characterized the trap up to temperatures of 300\,K, suggesting that room temperature operation is possible with this technology. % which can greatly reduce experimental footprints \cite{Spivey2021} and prove suitable for long-term scaling, is possible with this technology.

Because the traps were measured in a newly constructed cryogenic trapping apparatus for which noise had not previously been characterized \cite{Decaroli2021}, we discovered systematic noise consistent with an external source.
We further isolated this noise, identifying that the electrical configuration (primarily, grounding related to \ac{DC} voltage supplies) played a role, which indicates the need to improve source noise levels, external filtering, and grounding configuration.
Considering the remaining sources of noise, we found an axial mode heating rate dependence on mode frequency and trap temperature in accordance with models of surface effects like fluctuating patch potentials or electric field noise from dielectric bulk or surface layers.
We also observed the presence of a large stray static electric field, finding that it was stable and could be compensated.
Stray electric fields of this magnitude are sometimes linked to high heating rates \cite{Narayanan2011}, but we do not observe this. % maybe more explicitly what we see is, despite stray static electric fields comparable with many seen in the literature, we can achieve comparably low heating rates
% Overall, trap performance is well-suited for quantum computing and ion-based measurement applications. % YC et al.: just don't make these kinds of statements unless we show it. % See if we can find a way to restate the intro cleanly.

Our observations may also inform the design of the next generation of \ac{3D}-structured microfabricated ion traps.
For example, we observed that isotropically etched spacers produced sharp sidewall profiles, creating light scatter and possibly contributing to stray electric field that needed to be compensated.
% In the future, surface treatments might be applied or the spacer sidewalls flattened, recessed, or metal-coated, in order to further reduce stray fields.
These walls could be surface-treated, retracted, or flattened using different etching techniques (while ensuring mechanical stability) in order to reduce any limitation on heating rates due to surface dielectric effects. % this is kind of redundant, we mention it earlier
% The layer stack could be adjusted to reduce the capacitance and increase transition stability.
% Alternatively, the magnetic field gradients might be engineered to produce quantum gates \cite{Ospelkaus2011}. % or current-carrying electrodes with the possibility of generating large oscillating field gradients could be implemented to drive transitions electrically \cite{Ospelkaus2011}.
If noise sources originate in bulk dielectric such as these spacers, metal shielding layers could mitigate them.
Material and design engineering approaches, for example by thickening metal and dielectric layers, could lower ohmic and capacitive loss through the \ac{RF} electrodes, reducing power dissipation and lowering the temperature at which the trap can be operated.

The \ac{MEMS} microfabrication process used to construct the trap in this work is extensible, supporting a number of technologies that will manage growing complexity.
Connectivity between the trap and signal or measurement lines could be handled using through-substrate vias \cite{Guise2015}, while through-glass vias \cite{Shorey2016} could replace wirebond connections between wafers.
By integrating on-chip electronics \cite{Stuart2019}, one could increase signal density without increasing in-vacuum wiring requirements or burdening cryostats with excess heat loads.
Integrated waveguides \cite{Mehta2020, Niffenegger2020} would allow to channel light used for trapping, cooling, detection, and control to the ion, thus relaxing requirements for optical access. %and alignment of the wafers.
Junctions could be patterned along the trap axis \cite{Hensinger2006, Blakestad2009, Bowler2012, Shu2014, Decaroli2021a} to interface zones, though the spacer and top wafers must be designed carefully to allow optical access. % enable algorithms for splitting and interaction between logical qubit registers.
Through the use of vias and multiple metal layers, both top and bottom wafers support extensive electrode segmentation, which could extend to include the addition of \ac{RF} rails on either wafer. % , thus improving control over many ions.
Multiple trapping zones and \ac{RF} rails have been demonstrated using this technology in surface traps, including in a single-wafer predecessor of this trap \cite{Holz2020}.
Work to integrate several of these technologies into our microfabricated \ac{MEMS} trap is actively underway. 
% More extensive electrode segmentation on both bottom and top \ac{DC} electrodes here would improve control over many ions.
% If \ac{RF} signals were delivered to the top wafer, they could produce field configurations similar to macroscopic \ac{3D} traps. % , increasing trap depth.

Our demonstration of an industrially microfabricated trap with high performance, stable operation, and the potential to integrate optical and electronic features thus positions \ac{3D} \ac{MEMS} ion traps as a promising technology with which to underpin a scalable approach to trapped-ion quantum computing. % strong contenders to 
% underpins a promising approach towards scalable quantum computing with trapped ions.

\ack

% contributions
CR, EA, RB, TM, PS, and JH conceived the project. % and provided guidance along with PH.
SA, CD, CR, YC, and JH designed the trap; MV, CA, and SA performed trap simulations; SA led the fabrication with help of LP.
CA, CD, RO, and RM constructed the experimental apparatus; CA, CD, and SA performed trap characterization measurements and data analysis.
SA, CA, and JH wrote the manuscript with the help of all authors.

% funding
This work was supported by the EU H2020 FET Open project PIEDMONS (Grant No. 801285). We acknowledge support from IQI GmbH. 

% acks

% IFAT
We thank the Infineon Villach experts from Unit Process Development for setting up new processes, Unit Process Engineers for fabrication support, Failure Analysis for assistance in device characterization, Process Integration colleagues for fruitful discussions about MEMS technology, and Front-End operations for the execution of the ion trap fabrication. Moreover, we thank the Back-End experts at Infineon in Regensburg for module assembly.

% UIBK
We acknowledge G. Cerchiari for help with simulations.

% ETH
We further acknowledge L. Stolpmann for contributions to optical infrastructure and thank M. Marinelli and M. Stadler for assistance with control software.

\newpage

\appendix 

\section{Simulation methods}
\label{sec:methods_sim}

\subsection{Potential symmetry and anharmonicity}
\label{sec:anharmonicity}

In Section \ref{sec:concept} we investigated the effect of applying \ac{DC} voltages to increase confinement into the out-of-plane direction ($\hat{z}$) at the expense of confinement in the in-plane direction ($\hat{y}$).
The resulting voltage set solution, given in Table~\ref{tab:voltages}, produced the potential shown in Figure~\ref{fig:concept}c, which remains highly harmonic around the trapping site.
In particular, the ratio of the fourth-order term to the second-order term of an even polynomial fourth-order fit to the pseudopotential are 0.6\%, 0.7\%, and 8\% at trap coordinates $x=\SI{100}{\micro\meter}$, $y=\SI{50}{\micro\meter}$, and $z=\SI{50}{\micro\meter}$ relative to the trap center, respectively.
These distances are chosen because they (in the axial direction) represent the extent of a reasonable ion chain, and (in the radial directions) are on the order of the ion-electrode distance.
The effect of fourth-order anharmonicity sampled closer to the trap center will be smaller, since it scales with $x^2$.
These results are not significantly changed from the case with less out-of-plane confinement (the 0.2\,eV voltage set in Table~\ref{tab:voltages}), with ratios 0.6\%, 0.8\%, and 12\%.
From the change in out-of-plane asymmetry between the two voltage sets, it appears that the use of \ac{DC} to modify confinement improves the symmetry, which is to be expected since this produces a more quadrupole-like pseudopotential with larger effective trap efficiency.
From exploratory simulations, we see that of \ac{RF} drives on electrodes on the top wafer have a similar effect, but by modifying the intrinsic trap efficiency.

\section{Experimental methods}
\label{sec:methods}

\subsection{RF voltage calibration}
\label{sec:rfcal}

Values of $V_{\rm{rf}}$ were inferred primarily using a rectifier circuit that sampled a fraction of the \ac{RF} signal near the trap.
Using values of the circuit components specified to 1\% uncertainty, the rectifier signal was used to calibrate the temperature-dependent transfer function of the resonator used to step up the \ac{RF} voltage.
However, the pick-off fraction of the rectifier is also temperature dependent, increasing this uncertainty to about 5\%.

\subsection{Trap temperature calibration}
\label{sec:tempcal}

We found $T_{\rm{trap}}$ to be a good proxy for the temperature of the bottom wafer trap surface (on which the \ac{RF} electrodes were present) $T_{\rm{surface}}$ when the thermistor was also used as a heater over the range 75--300\,K. % (because heat could be dissipated faster through the thermal connection to the cooling stage than to the trap surface) % to obtain a ``cooldown calibration'', 
We used a temperature-dependent signature of the complex-valued \ac{RF} reflection signal $\tilde{S}_{11}$ to calibrate $T_{\rm{surface}}$, and then relate this to $T_{\rm{trap}}$.
The \ac{RF} signal is delivered to the electrodes via a coaxial transmission line, through a step-up resonator, and then through intermediate \ac{PCB}s onto the trap chip.
We want to access the portion of the $\tilde{S}_{11}$ signature that depends only a change in temperature of the trap chip.

First, we slowly cooled the cryostat from 300\,K to 7\,K, with no current flowing through the Pt1000 thermistor and no constant drive applied to the \ac{RF} electrodes, and measured the temperature near the \ac{RF} resonator $T_{\rm{RF}}$, the temperature on the intermediate \ac{PCB}s $T_{\rm{PCB}}$, and $\tilde{S}_{11}$.
Under these conditions, the temperatures of these thermometers --- all at the cryostat base temperature stage --- were assumed to be in equilibrium, including with $T_{\rm{surface}}$ and $T_{\rm{trap}}$.
(This also served to calibrate $T_{\rm{trap}}$.)
Further, our \ac{RF} resonator is constructed using internal coils soldered to a outer body, and so we expect the whole structure to be well thermally connected at a temperature well represented by $T_{\rm{RF}}$.
Therefore we know the temperatures of the different elements in the \ac{RF} circuit that could each contribute to $\tilde{S}_{11}$.

% We compare this temperature at the base of the trap with the temperature at the trap surface,  as inferred from RF reflection characteristics.
% To calibrate $T_{\rm{surface}}$, we first measured the complex-valued \ac{RF} reflection signal $\tilde{S}_{11}$ and the temperatures $T_{\rm{trap}}$ and $T_{\rm{RF}}$ (from a base-stage thermometer located near the stepped-up output of the \ac{RF} resonator) 
 % while measuring a calibrated thermometer placed next to the \ac{RF} resonator, as well as the .
Once the cooldown calibration was complete, and cryostat had settled at a base temperature, we used the trap thermistor to raise $T_{\rm{trap}}$ over a range 7--300\,K.
% Over this large range in $T_{\rm{trap}}$ achieved by heating,
Correspondingly, $T_{\rm{RF}}$ and $T_{\rm{PCB}}$ only increased from \SI{7}{\kelvin} to below \SI{8}{\kelvin}, suggesting that changes in the \ac{RF} impedance of the trap, rather than the resonator or other parts of the circuit, were primarily responsible for the behavior of $\tilde{S}_{11}$.
The center frequency and width of the resonance were extracted from $\tilde{S}_{11}$ over the range of the cooldown calibration, which vary monotonically with $T_{\rm{RF}}$, to be used as a signature with which to infer $T_{\rm{surface}}$.
Finally, while heating the thermistor, the inferred surface temperature $T_{\rm{surface}}$ was seen to agree with $T_{\rm{trap}}$ to within \SI{~5}{\kelvin}, suggesting that $T_{\rm{trap}}$ and is a reasonable measurement of the temperature of the bottom wafer trap surface, $T_{\rm{surface}}$, under steady-state operating conditions.
Since a constant \ac{RF} drive during trap operation does not easily permit the probe measurements of $\tilde{S}_{11}$ to infer $T_{\rm{surface}}$, its close relation to $T_{\rm{trap}}$ provides a helpful proxy measurement.
Though we do not know how well $T_{\rm{surface}}$ represents the temperature of all surfaces, this parameter is still useful for characterizing trap behavior relative to the temperature at the trap base, $T_{\rm{trap}}$, as seen in Section \ref{sec:hr}.

\subsection{Electrical configuration} % add schematic if deemed necessary before submission or from reviewer comments
\label{sec:elecconfig}

In early measurements of trap \#1, we had found that the grounding configuration had a significant effect on heating rates.
There, only four signal lines connected the trap ground (at the base cooling stage) to the signal source ground (external to the cryostat) through a total resistance $R \sim \SI{1}{\ohm}$, and measured heating rates exceeded $10^3$\,phonons/s.
Only after also connecting the signal source ground to the trap via the cryostat mechanical chassis, lowering the \ac{DC} resistance of this path to an estimated \SI{1}{\milli\ohm} (set by the contact resistance of cooling stage components), did we measure the heating rates reported in this work. % https://www.sciencedirect.com/science/article/abs/pii/S0011227519300943
These configuration changes were similar to those reported in ref. \cite{Decaroli2021a}.

In trap \#2, we varied the \ac{DC} electrical configuration further: bypassing external filters and amplifiers, disconnecting thermometers and heaters, and removing voltage source and ground line connections.
% Removing external filtering of \ac{DC} signals when the trap was connected produced no change in heating rate, nor did disconnecting possible noise transmitters like thermometers and heaters.
Heating rates were measured before, after, and upon reversal of each change to confirm that the measured effects were causal.
These changes produced the ``\ac{DC} disconnected'' results given in Section \ref{sec:hr}.

\section{Data analysis methods}
\label{sec:analysis}

\subsection{Heating rate normalization}
\label{sec:normalization}

Heating rates measured at frequencies $\{f_{m}\}$ while sweeping the trap temperature in Figure~\ref{fig:hr}b were frequency-normalized to frequencies $\{f_{\text{norm},m}\}$ by multiplying them by a factor
\begin{equation}
    \chi_f = \left(\frac{f_{\text{norm},m}}{f_{m}}\right)^{-\alpha_m}
\end{equation}
using the frequency exponents $\{\alpha_m\}$ from the fits in Figure~\ref{fig:hr}a, where $m$ indexes the motional modes.
While the heating rate of the ion is the result of contributions from all sources, our simple normalization method assume and corrects for a single limiting source.
We found exponents $(-2.3(1), -3.4(3), -2.9(3))$ for axial, in-plane radial, and out-of-plane radial modes.
The normalized frequencies $\{f_{\text{norm},m}\}$ were chosen to be the mean frequency values of modes in the data set, $(0.98, 2.35, 2.56)$\,MHz.
Frequencies here were closely clustered, with standard deviations $(0.08, 0.35, 0.29)$\,MHz, giving average normalization factors $(1.01, 1.09, 1.09)$.
Therefore, the scaling exponent $\beta$, and thus the temperature normalization of Figure~\ref{fig:hr}a data, are not subject to significant additional uncertainty due to the frequency normalization of Figure~\ref{fig:hr}b data.

Normalization of the data in Figure~\ref{fig:hr}c follows the same procedure.
These data have a relatively small spread in average frequencies $(2.92, 3.05)$\,MHz, with standard deviations both around \SI{30}{\kilo\hertz}.
When normalized to the the mean frequency of both modes' data (2.99\,MHz), this results in average normalization factors $(0.90, 1.08)$).

% We chose not to normalize the data presented in Figure \ref{fig:hr}a.
In Figure~\ref{fig:hr}a, the axial mode data was taken at a constant temperature $T_{\rm{trap}} = 185$\,K and did not require temperature normalization.
The radial mode data required that the \ac{RF} voltage was varied, which changed trap temperature between 75\,K and 193\,K (with standard deviations both around 34\,K).
Thus we considered the effect of a normalization to $T_{\rm{norm}} = \SI{185}{\kelvin}$, wherein radial heating rates would be multiplied by a factor
\begin{equation}
    \chi_T = \left(\frac{T_{\rm{norm}}}{T_{\rm{trap}}}\right)^{\beta}
\end{equation}
using the temperature scaling exponents 0.5(2) and 0.8(1)) for in-plane and out-of-plane radial modes, respectively.
We found that normalization did not significantly change the behavior of the radial mode data in Figure \ref{fig:hr}a, with average normalization factors $(1.32, 1.20)$.
% From the observations discussed in the text, the temperature normalization of axial data likely corrects for surface-based effects, while
Radial modes are likely dominated by noise external to the trap, which should be largely independent of trap temperature.
With this physical understanding and the weak correlation of radial heating rates with an exponential temperature scaling model, we choose not to apply this normalization when displaying the data in Figure \ref{fig:hr}a.
% Radial normalization corrects for the dependence of external noises on trap temperature, 
% As expected for noise sources external to the trap, changing trap temperature has little effect,
% We chose to normalize all modes using this exponent because it represents the scaling behavior most consistent with a surface-based noise source, the aspect of noise for which this normalization corrects.
% We assume that all modes' heating rates will scale in a manner that follows the axial temperature scaling.

\subsection{Uncertainty of repeated heating rate measurements}
\label{sec:uncertain_hr}

In Section \ref{sec:hr} we investigated the temporal variation of heating rates to understand possible noise sources.
We sought to understand whether, in the presence of noise assumed to produce the significant scatter of radial mode heating rates relative to frequency and temperature, individual measurement uncertainties were representative of the expected distribution.
Varying noise could lead to measured values appearing beyond the normal distribution implied by a single measurement's uncertainty.
If the noise changed its characteristics over timescales longer than a single measurement, but before a complete data set could be acquired, this could have consequences for our interpretation of data set trends.

Nineteen sequential measurements were performed over nearly 8~hours and under identical measurement conditions, and gave values 43(9), 33(23), 30(18), 10(28), 50(17), 45(15), 34(17), 67(20), 31(16), 36(25), 9(14), 42(25), 18(18), 29(23), 16(25), 69(27), 91(20), 78(23), 37(25)\,phonons/s for the axial mode, 97(5), 115(10), 113(10), 109(10), 115(8), 94(8), 110(5), 96(6), 107(7), 109(7), 126(9), 109(9), 120(8), 94(10), 96(7), 103(6), 99(12), 100(9), 117(10)\,phonons/s for the out-of-plane radial mode, and 45(9), 15(4), 16(3), 22(3), 19(3), 21(3), 17(3), 21(3), 20(3), 16(3), 24(3), 20(3), 18(3), 20(3), 11(3), 15(3), 23(3), 23(4), 12(4)\,phonons/s for the in-plane radial mode.
The values in parentheses represent the standard error associated with heating rate fits for individual measurements in the data set (at specific times).
The data sets have mean values $41$, $106$, and $20$\,phonons/s, while the standard deviations are $22$, $9$, and $7$\,phonons/s, which does not take into account the measurement uncertainties.
The mean values of the uncertainties within each set are 20, 8.1, and 3.5\,phonons/s, which are close the sets' standard deviations.
It is thus apparent that while some changes can cause individual heating rates to vary over short timescales, reported measurement uncertainties accurately represent the distribution of measurement values.

We found similar results in the case where radial shim fields were applied to test the robustness of measured rates to shifts in micromotion.
Here, we measured 36(34), 65(26), 19(21), 24(20), 43(16), 56(20), 62(17)\,phonons/s for the axial mode, 96(7), 112(7), 115(10), 113(9), 106(10), 106(11), 109(11)\,phonons/s for the out-of-plane radial mode, and 17(2), 37(17), 29(4), 19(4), 17(5), 57(19), 21(3)\,phonons/s for the in-plane radial mode.
The respective mean values are $44$, $108$, and $28$\,phonons/s, with standard deviations $17$, $6$, and $14$\,phonons/s.
The mean values of the uncertainties here are 22, 9.4, and 7.7\,phonons/s, which are comparable to the set's standard deviations.
Therefore the possibility of changing noise characteristics over the course of experimental data acquisition is not a significant concern.

\newpage

%%%%%%%%%%%%%%
% References %
%%%%%%%%%%%%%%

\bibliographystyle{MSP}
\bibliography{references}

\end{document}